\documentclass{iopart}
\usepackage{euscript,iopams,amssymb,amsfonts,graphicx,bm}
\usepackage{pgfplots,setstack}

\usepackage{float}
\usepackage{epsfig}
\usepackage{esint}
\usepackage{color}

\usepackage{bm,braket}

\eqnobysec

\newcommand{\q}{\widetilde{q}}

\newcommand{\p}{\widetilde{p}}

\newcommand{\Markov}[2]{\underset{#1}{\overset{#2}{\rightleftharpoons}}}
\newcommand{\R}{{\mathbb R}}
\newcommand{\C}{\widetilde{C}}

\newcommand{\calU}{{\mathcal U}}
\newcommand{\calT}{{\mathcal T}}

\renewcommand{\e}{\mathrm{e}}

\newcommand{\x}{\mathbf{x}}

\newcommand{\X}{\mathbf{X}}
\renewcommand{\P}{\mathbb{P}}

\newcommand{\n}{\mathbf n}

\newcommand{\E}{{\mathbb E}}

\begin{document}
\title[Search processes with partially absorbing targets]{Search processes with stochastic resetting and partially absorbing targets}
\author{Ryan D. Schumm and Paul C. Bressloff}
\address{Department of Mathematics, University of Utah, Salt Lake City, UT, USA} \ead{bressloff@math.utah.edu}

\begin{abstract}
We extend the theoretical framework used to study search processes with stochastic resetting to the case of partially absorbing targets. Instead of an absorption event occurring when the search particle reaches the boundary of a target, the particle can diffuse freely in and out of the target region and is absorbed at a rate $\kappa$ when inside the target. In the context of cell biology, the target could represent a chemically reactive substrate within a cell or a region where a particle can be offloaded onto a nearby compartment. We apply this framework to a partially absorbing interval and  to spherically symmetric targets in $\R^d$. In each case, we determine how the mean first passage time (MFPT) for absorption depends on $\kappa$, the resetting rate $r$, and the target geometry. For the given examples, we find that the MFPT is a monotonically decreasing function of $\kappa$, whereas it is a unimodal function of $r$ with a unique minimum at an optimal resetting rate $r_{\rm opt}$. The variation of $r_{\rm opt}$ with $\kappa$ depends on the spatial dimension $d$, decreasing in sensitivity as $d$ increases. For finite $\kappa$, $ r_{\rm opt}$ is a non-trivial function of the target size and distance between the target and the reset point. We also show how our results converge to those obtained previously for problems with totally absorbing targets and similar geometries when the absorption rate becomes infinite. Finally, we generalize the theory to take into account an extended chemical reaction scheme within a target.
\end{abstract}

\maketitle

\section{Introduction}

\begin{figure}[b!]
\raggedleft
\includegraphics[width=12cm]{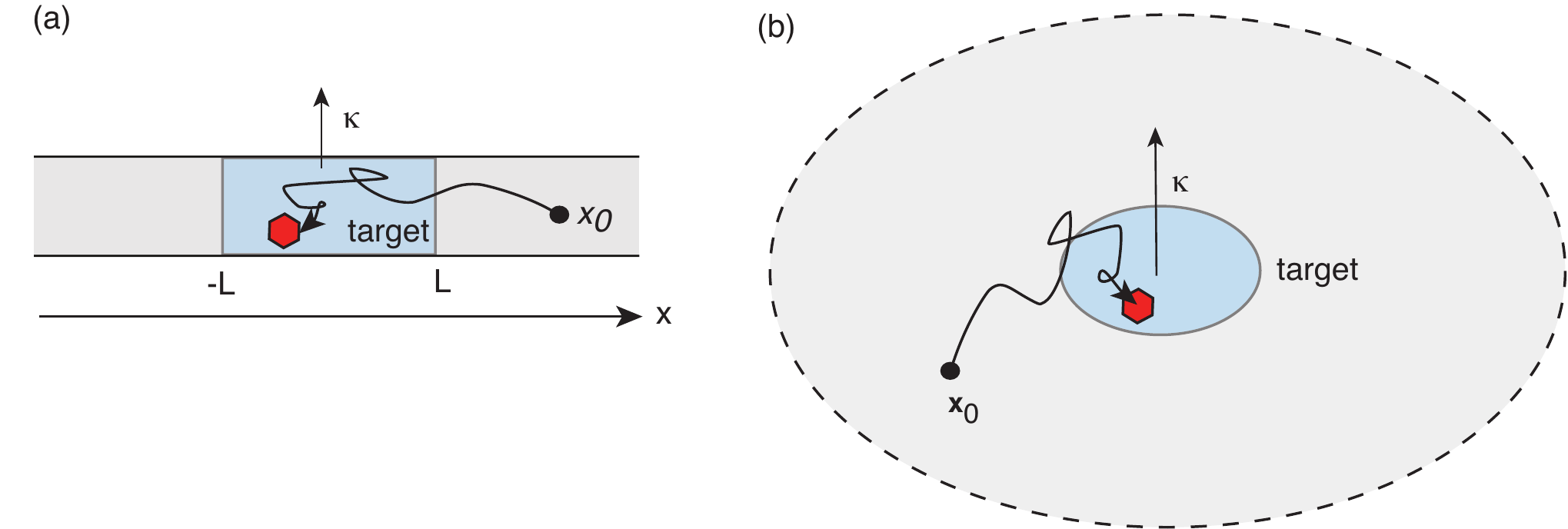} 
\caption{A partially absorbing target in (a) 1D and (b) 2D }
\label{fig1}
\end{figure}

In studies of random search processes, targets are either located on the boundary of the search domain or in the interior. In the latter case, one is typically interested in solving the first passage time (FPT) problem to reach the target boundary, which is usually taken to be totally absorbing. One could also consider a partially absorbing boundary, whereby there is a non-zero probability that the particle is reflected rather than absorbed by the target. However, there are a growing number of examples in cell biology where the whole interior target domain $\calU$ acts as a partial absorber rather than the target boundary $\partial \calU$ \cite{Bressloff13}, as illustrated in Fig. \ref{fig1}. Now the particle can freely enter and exit $\calU$, and is absorbed at a rate $\kappa$ when inside $\calU$. One example is the motor-driven intracellular transport of vesicles to en passant synaptic targets in the axons and dendrites of neurons \cite{Bressloff20b}. The particle (searcher) represents a molecular motor complex moving along a microtubule, which is modeled as a 1D random search process. The transport is often bidirectional, with the particle randomly switching between anterograde and retrograde motion according to a velocity-jump process. In the fast switching limit, the stochastic dynamics can be approximated by diffusive-like search \cite{Newby10}. In this application, absorption corresponds to the transfer of the vesicular cargo to a synaptic target, say. Another example of a partially absorbing target is a local chemical substrate within a cell. This follows from the observation that random search processes are also used to model diffusion-limited reactions, in which chemical species A has to diffuse within range of chemical species B, say, in order to react \cite{Rice85}.  The transport process can be formulated as a searcher A looking for a target B, and the effective reaction rate related to the mean FPT (MFPT). In the case of a diffusion-limited reaction, as soon as the particle hits the boundary of the reaction domain or target it reacts instantaneously. However, it is also possible that the target represents a spatially extended chemical substrate $\calU$ rather than a single molecule B, such that particle A reacts at some rate $\kappa$ when within $\calU$. The target domain $\calU$ is then partially absorbing.

Passive diffusive search often results in unrealistically slow reaction rates. One mechanism for significantly reducing the search time is random intermittent transport, in which a reactant alternates between periods of passive diffusion and active ballistic motion \cite{Loverdo08,Benichou10,Benichou11}; the active phase could be mediated by molecular motors transiently binding diffusing reactant molecules within the cellular environment. In many cases one finds that the mean search time can be minimized by varying the rates of switching between the different transport states. A simplified version of a random intermittent search process is diffusion with stochastic resetting or restart \cite{Evans11a,Evans11b,Evans14}. The position of a Brownian particle (reactant) is reset to a fixed location at a random sequence of times that is generated by a Poisson process, say, and the mean first passage time (MFPT) to find a target is minimized at an optimal resetting rate (assuming it exists). Although resetting was originally formulated in terms of Brownian motion, it has subsequently been applied to wide range of stochastic search processes, see the recent review \cite{Evans20} and references therein. 

In this paper we develop the basic theory of diffusive search processes with stochastic resetting and partially absorbing targets. We first consider a single partially absorbing target (section 2). We calculate the MFPT and determine how it depends on the absorption rate $\kappa$, the resetting rate $r$, and the target geometry. We then apply the theory to a partially absorbing interval (section 3) and a spherically symmetric target in $\R^d$ (section 4). In each case we find that the MFPT is a monotonically decreasing function of $\kappa$, whereas it is a unimodal function of $r$ with a unique minimum at an optimal resetting rate $r_{\rm opt}$. The variation of $r_{\rm opt}$ with $\kappa$ depends on the spatial dimension $d$, decreasing in sensitivity as $d$ increases. For finite $\kappa$, $ r_{\rm opt}$ is a non-trivial function of the target size and distance between the target and the reset point. We also show how our results converge to those obtained previously for problems with totally absorbing targets and similar geometries when the absorption rate becomes infinite. We then generalize the theory to the case of multiple partially absorbing targets (section 5) and to extended chemical reactions schemes (section 6).

\setcounter{equation}{0}
\section{Diffusive search for a single partially absorbing target in $\R^d$}
Consider a partially absorbing target $\calU_1\subset \R^d$. 
Let $p(\x,t|\x_0)$ be the probability density that at time $t$ a particle is at $\X(t)=\x$, having started at position $\x_0$. We will set $p=q$ for all $\x\in \R^d\backslash \calU_1$ and $p=p_1$ for all $\x\in \calU_1$ such that 
\numparts 
\begin{eqnarray}
\label{mastera}
	\frac{\partial q(\x,t|\x_0)}{\partial t} &=& D\nabla^2 q(\x,t|\x_0), \ \x\in \R^d \backslash \calU_1,\\
	\label{masterb}
	\frac{\partial p_1(\x,t|\x_0)}{\partial t} &=& D\nabla^2 p_1(\x,t|\x_0) -\kappa p_1(\x,t|\x_0),\ \x\in \calU_1,
	\end{eqnarray}
together with the continuity conditions 
\begin{equation}
\label{masterc}
\fl q(\x,t|\x_0)=p_1(\x,t|\x_0),\quad \nabla q(\x,t|\x_0)\cdot \n =  \nabla p_1(\x,t|\x_0)\cdot \n,\quad \x \in \partial \calU_1,
\end{equation}
\endnumparts
and the initial conditions $q(\x,t|\x_0)=\delta(\x-\x_0)$ and $p_1(\x,0|\x_0)=0$.
Here $\kappa$ is the rate at which the particle is absorbed by the target and $\n$ is the unit normal to the surface of the target. Since we take the particle to start outside the domain $\calU_1$, it follows that in the limit $\kappa \rightarrow \infty$ the particle is immediately absorbed as soon as it hits the target boundary. Hence, we recover a totally absorbing target with $q$ evolving according to equation (\ref{mastera}) and
\begin{equation}
q(\x,t|\x_0)=0,\quad \x \in \partial \calU.
\end{equation}
Note that our definition of a partially absorbing target for finite $\kappa$ is different from the definition of a target with a partially absorbing boundary condition
\begin{equation}
\label{master0}
\nabla q\cdot \n+\alpha  q=0,\ \x \in \partial \calU_1
\end{equation}
for some $\alpha \geq 0$. In the latter case, partial absorption means that there is a non-zero probability that the particle is reflected at the boundary.

The probability flux for absorption at time $t$ is 
\begin{eqnarray}
\label{J}
	J(\x_0,t)&= \kappa\int_{\calU_k} p_1(\x,t|\x_0)d\x.
	\end{eqnarray}
Hence, the probability that the particle is eventually absorbed by the target is
\begin{equation}
\label{split}
\pi(\x_0)=\int_0^{\infty}J(\x_0,t')dt' =\widetilde{J}(\x_0,0),
\end{equation}
where $\widetilde{J}(\x_0,s)$ denotes the Laplace transform of $J(\x_0,t)$.
Next we introduce the survival probability that the particle hasn't been absorbed by the target in the time interval $[0,t]$, having started at $\x_0$:
\begin{equation}
\label{Q1}
\fl Q(\x_0,t)=\int_{\R^d}p(\x,t|\x_0)d\x=\int_{\R^d\backslash \calU_1}q(\x,t|\x_0)d\x+ \int_{\calU_1}p_1(\x,t|\x_0)d\x.
\end{equation}
Differentiating both sides of this equation with respect to $t$ and using equations (\ref{mastera}) and (\ref{masterb}) implies that
\begin{eqnarray}
\fl \frac{\partial Q(\x_0,t)}{\partial t}&=D\int_{\R^d\backslash \calU_1}\nabla\cdot \nabla q(\x,t|\x_0)d\x\nonumber \\
\fl &\quad +\int_{\calU_1}\left [D\nabla\cdot \nabla p_1(\x,t|\x_0)-\kappa p_1(\x,t|\x_0)\right ]d\x\nonumber \\
\fl &= - D\int_{\partial \calU_1}\nabla q\cdot \n d\sigma+ D\int_{\partial \calU_1}\nabla p_1\cdot \n d\sigma -\kappa  \int_{ \calU_1}p_1(\x,t|\x_0)d\x\nonumber \\
\fl &=-\kappa  \int_{ \calU_1}p_1(\x,t|\x_0)d\x,
\label{Q2}
\end{eqnarray}
where we have used the current conservation condition in equation (\ref{masterc}).
Laplace transforming equation (\ref{Q2}) and imposing the initial condition $Q(\x_0,0)=1$ gives
\begin{equation}
\label{QL}
s\widetilde{Q}(\x_0,s)-1=- \widetilde{J}(\x_0,s).
\end{equation}
It immediately follows from equation (\ref{split}) that
\begin{equation}
\label{split0}
\pi(\x_0)=1 -\lim_{s\rightarrow 0}s\widetilde{Q}(\x_0,s)=1-Q_{\infty}(\x_0).
\end{equation}
Here $1-Q_{\infty}(\x_0)$ is the probability that the particle is eventually absorbed by the target. It is well known that Brownian motion in 1D and 2D is recurrent so that $\pi(\x_0)=1$, whereas it is transient in 3D, $\pi_1(\x_0)<1$. In all cases the MFPT is infinite when the search domain is unbounded.

Now suppose that prior to being absorbed by one of the targets, the particle can instantaneously reset to a fixed location $\x_r$ at a random sequence of times generated by an exponential probability density $\psi(\tau)=r\e^{-r\tau}$, where $r$ is the resetting rate. The probability that no resetting has occurred up to time $\tau$ is then $\Psi(\tau)=1-\int_0^{\tau}\psi(s)ds=\e^{-r\tau}$. In the following we identify $\x_r$ with the initial position by setting $\x_0=\x_r$. (One could also incorporate resetting delays -- finite return times and refractory periods - and non-exponential resetting statistics \cite{Mendez19,Evans19a,Mendez19a,Bodrova20,Pal20,Bressloff20A,Evans20}. However, we focus on the simplest case here.)
Equations (\ref{mastera}) and (\ref{masterb}) become
\numparts 
\begin{eqnarray}
\label{mastera(res)}
\fl	\frac{\partial q_r(\x,t|\x_0)}{\partial t} &=& D\nabla^2 q(\x,t|\x_0)-rq_r(\x,t|\x_0)+r\delta(\x-\x_0), \ \x\in \R^d \backslash \calU_1,\\
	\label{masterb(res)}
\fl	\frac{\partial p_{r,1}(\x,t|\x_0)}{\partial t} &=& D\nabla^2 p_{r,1}(\x,t|\x_0) -(\kappa+r) p_{r,1}(\x,t|\x_0),\ \x\in \calU_1,
	\end{eqnarray}
	\endnumparts
Let $Q_r(\x_0,t)$ be the corresponding survival probability in the presence of resetting:
\begin{equation}
\label{surv}
\fl Q_r(\x_0,t)=\int_{\R^d}p_r(\x,t|\x_0)d\x=\int_{\R^d\backslash \calU_1}q_r(\x,t|\x_0)d\x+ \int_{\calU_1}p_{r,1}(\x,t|\x_0)d\x.
\end{equation} 
One now observes that $Q_r$ can be related to the survival probability without resetting, $Q$, using a last renewal equation \cite{Evans11a,Evans11b,Evans20}; this holds irrespective of whether the target is partially or totally absorbing:
\begin{equation}
\label{renQ}
{Q_r(\x_0,t)=\e^{-rt}Q(\x_0,t)+r\int_0^tQ(\x_0,\tau)Q_r(\x_0,t-\tau)\e^{-r\tau}d\tau.}
\end{equation}
The first term on the right-hand side represents trajectories with no resettings. The integrand in the second term is the contribution from trajectories that last reset at time $\tau\in (0,t)$, and consists of the product of the survival probability starting from $\x_0$ with resetting up to time $t-\tau$ and the survival probability starting from $\x_0$ without any resetting for the time interval $\tau$. Since we have a convolution, it is natural to introduce the Laplace transform
\[ \widetilde{Q}_r(\x_0,s)=\int_0^{\infty}Q_r(\x_0,t)\e^{-st}dt.\]
Laplace transforming the last renewal equation and rearranging shows that
\begin{equation}
\label{Qr}
{ \widetilde{Q}_r(\x_0,s)=\frac{ \widetilde{Q}(\x_0,r+s)}{1-r \widetilde{Q}(\x_0,r+s)}.}
 \end{equation}

A well known property of diffusive search in unbounded domains with stochastic resetting is that the probability of finding the target is now unity, $\pi_{r}(\x_0)=1$, and the MFPT is rendered finite, $T_r(\x_0)<\infty$.
The MFPT can be expressed in terms of $Q_r$ according to
\begin{eqnarray}
\label{TQ}
T_r(\x_0)&=- \int_0^{\infty}t \frac{dQ_r(\x_0,t)}{dt}d\tau=\int_0^{\infty}Q_r(\x_0,t)dt,
\end{eqnarray}
This follows from the fact that the FPT density $f_r(\x_0,t)$ is related to the survival probability according to
\begin{equation}
\label{fr}
f_r(\x_0,t)=-  \frac{dQ_r(\x_0,t)}{dt}.
\end{equation}
Combining equations (\ref{Qr}) and (\ref{TQ}) shows that
the MFPT to reach the target is then given by
\begin{equation}
\label{Ttot}
T_r(\x_0) =\widetilde{Q}_r(\x_0,0)=\frac{ \widetilde{Q}(\x_0,r)}{1-r \widetilde{Q}(\x_r,r)}=\frac{1-  \widetilde{J}(\x_0,r)}
{  r\widetilde{J}(\x_0,r)}.
 \end{equation}
 Equation (\ref{Ttot}) implies that the MFPT with resetting is determined by the Laplace transform of the flux, $\widetilde{J}(\x_0,r)$, which depends on the absorption rate $\kappa$. In order to determine the flux we thus have to solve the Laplace transformed version of equations (\ref{mastera})--(\ref{masterc}), which take the form
 \numparts 
\begin{eqnarray}
\label{masterLTa}
	&D\nabla^2 \q(\x,s|\x_0)-s\q(\x,s|\x_0)=-\delta(\x-\x_0), \ \x\in \R^d \backslash \calU_1,\\
	\label{masterLTb}
	&D\nabla^2 \p_1(\x,s|\x_0) -(\kappa+s) \p_1(\x,s|\x_0)=0,\ \x\in \calU_1,
	\end{eqnarray}
together with the continuity conditions 
\begin{equation}
\label{masterLTc}
\fl \q(\x,s|\x_0)=\p_1(\x,s|\x_0),\quad \nabla \q(\x,s|\x_0)\cdot \n =  \nabla \p_1(\x,s|\x_0)\cdot \n\quad \x \in \partial \calU_1.
\end{equation}
	\endnumparts

\setcounter{equation}{0}
\section{Partially absorbing interval in $\R$}
As our first, example, suppose that there is a single target $\calU_1=[-L,L]$ in $\R$, see Fig. \ref{fig1}(a). Equations (\ref{masterLTa})-(\ref{masterLTc}) become
\numparts 
\begin{eqnarray}
\label{master1Da}
	\frac{\partial q_-(x,t|x_0)}{\partial t} &= D\frac{\partial^2 q_-(x,t|x_0)}{\partial x^2}, \ x < L,	\\
	\label{master1Db}
	\frac{\partial p_1(x,t|x_0)}{\partial t} &= D\frac{\partial^2 p_1(x,t|x_0)}{\partial x^2} -\kappa p_1(x,t|x_0),\ -L<x<L,\\
	\label{master1Dc}
	\frac{\partial q_+(x,t|x_0)}{\partial t} &= D\frac{\partial^2 q_+(x,t|x_0)}{\partial x^2}, \ x > L,
		\end{eqnarray}
		\endnumparts
together with the matching conditions 
\numparts
\begin{eqnarray}
q_-(-L,t|x_0)&=p_1(-L,t|x_0),\ \frac{\partial q_-}{\partial x}(-L,t|x_0)  = \frac{\partial p_1}{\partial x}(-L,t|x_0),\\
q_+(L,t|x_0)&=p_1(L,t|x_0),\ \frac{\partial q_+}{\partial x}(L,t|x_0)  = \frac{\partial p_1}{\partial x}(L,t|x_0).
\end{eqnarray}
\endnumparts
Take the initial conditions $q_{+}(x,0|x_0)=\delta(x-x_0)$ and $p(x,0|x_0)=0=q_-(x,0|x_0)$. In other words, $x_0>L$. The corresponding equations in Laplace space are of the form
\numparts 
\begin{eqnarray}
\label{master1DLTa}
	&  D\frac{\partial^2 \q_-(x,s|x_0)}{\partial x^2}-s \q_-(x,t|x_0)=0, \ x < L,	\\
	  \label{master1DLTb}
	&D\frac{\partial^2 \p_1(x,s|x_0)}{\partial x^2}  -(s+\kappa) \p_1(x,s|x_0)= 0,\ -L<x<L,\\
	\label{master1DLTc}
	& D\frac{\partial^2 \q_+(x,s|x_0)}{\partial x^2}-s \q_+(x,t|x_0)=-\delta(x-x_0), \ x > L,
		\end{eqnarray}
\endnumparts
together with the same matching conditions as before.

The general bounded solutions of equations (\ref{master1DLTa}) and (\ref{master1DLTb}) are
\begin{eqnarray}
\fl    \q_-(x, s|x_0) = A(s)e^{\alpha(s) x},\quad  \p_1(x, s|x_0) = C(s)e^{\beta(s) x} + E(s)e^{-\beta(s) x},
\end{eqnarray}
where \begin{eqnarray} 
\alpha(s) = \sqrt{s/D},\quad \beta(s) = \sqrt{(s + \kappa)/D}.
\end{eqnarray}
Note that the coefficients $A,C, E$ also depend on the initial position $x_0$.  For equation (\ref{master1DLTc}), we can write the solution in the form
\begin{eqnarray}
    \q_+(x, s|x_0) = \q_{+, h}(x, s) + G(x, s; x_0),
\end{eqnarray}
where $\q_{+, h}$ is the solution to the homogeneous problem that satisfies the appropriate non-homogeneous boundary conditions and $G$ is the $1$-D Green's function that satisfies equation (\ref{master1DLTc}) and the homogeneous boundary conditions at $x = L$ and $x = \infty$. The Green's function can be written as
\begin{eqnarray}
  \fl   G(x, s; x_0) 
    &= -\frac{e^{2\alpha(s) L}}{2\alpha(s) D}\left[\Theta(x_0 - x)\phi_1(x, s)\phi_2(x_0, s) + \Theta(x - x_0)\phi_1(x_0, s)\phi_2(x, s)\right],
\end{eqnarray}
where $\Theta$ is the Heaviside function and
\begin{eqnarray}
    \phi_1(x, s) = e^{-\alpha(s) x} - e^{\alpha(s) (x - 2L)} \quad \makebox{and} \quad \phi_2(x, s) = e^{-\alpha(s) x}. 
\end{eqnarray}
The homogeneous solution is given by
\begin{eqnarray}
    \q_{+, h}(x, s|x_0) = F(s)e^{-\alpha(s)(x - L)},
\end{eqnarray}
where $F(s) = \p(L, s|x_0)$.  Applying the matching conditions and solving for $A(s)$, $C(s)$, $E(s)$, and $F(s)$ yields
\numparts
\begin{eqnarray}
  &A(s) = \frac{\beta(s)}{2D\Phi}e^{\alpha(s)(x_0 - 2L)},\\
    &C(s) = \frac{(\alpha(s) + \beta(s))}{4D\Phi(s)}e^{-\alpha(s)(x_0 - L)}e^{\beta(s) L},\\
      &E(s) = \frac{(\beta(s) - \alpha(s))}{4D\Phi(s)}e^{-\alpha(s)(x_0 - L)}e^{-\beta(s) L},\\
     &F(s)= \frac{e^{-\alpha(s)(x_0 - L)}}{2D\Phi(s)}\left[\alpha(s)\sinh{(2\beta(s) L)} + \beta(s)\cosh{(2\beta(s) L)}\right]
\end{eqnarray}
\endnumparts
with
\begin{eqnarray}
 \Phi(s)& = \left[\alpha(s)\sinh{(\beta(s) L)} + \beta(s)\cosh{(\beta(s) L)}\right]\nonumber \\
 &\quad \times \left[\alpha(s)\cosh{(\beta(s) L)} + \beta(s)\sinh{(\beta(s) L)}\right].
\end{eqnarray}
By equation (\ref{split}), we can write the hitting probability as
\begin{eqnarray}
  \pi(x_0) 
  &= \kappa\int_{-L}^L\p_1(x,0|x_0)dx= \kappa\int_{-L}^L \left(C(0)e^{\beta(0) x} + E(0)e^{-\beta(0) x}\right)dx\nonumber \\
 &= \frac{2\kappa}{\beta(0)}(C(0) + E(0))\sinh{(\beta(0) L)}.
\end{eqnarray}
Since
\begin{eqnarray*}
    C(0) + E(0) = \frac{1}{2D\beta(0)\sinh{(\beta(0)L)}}.
\end{eqnarray*}
we see that $\pi_0(x_0) = 1$, which is expected since one-dimensional diffusion is recurrent.  

\begin{figure}[t!]
    \raggedleft
    \includegraphics[width=13cm]{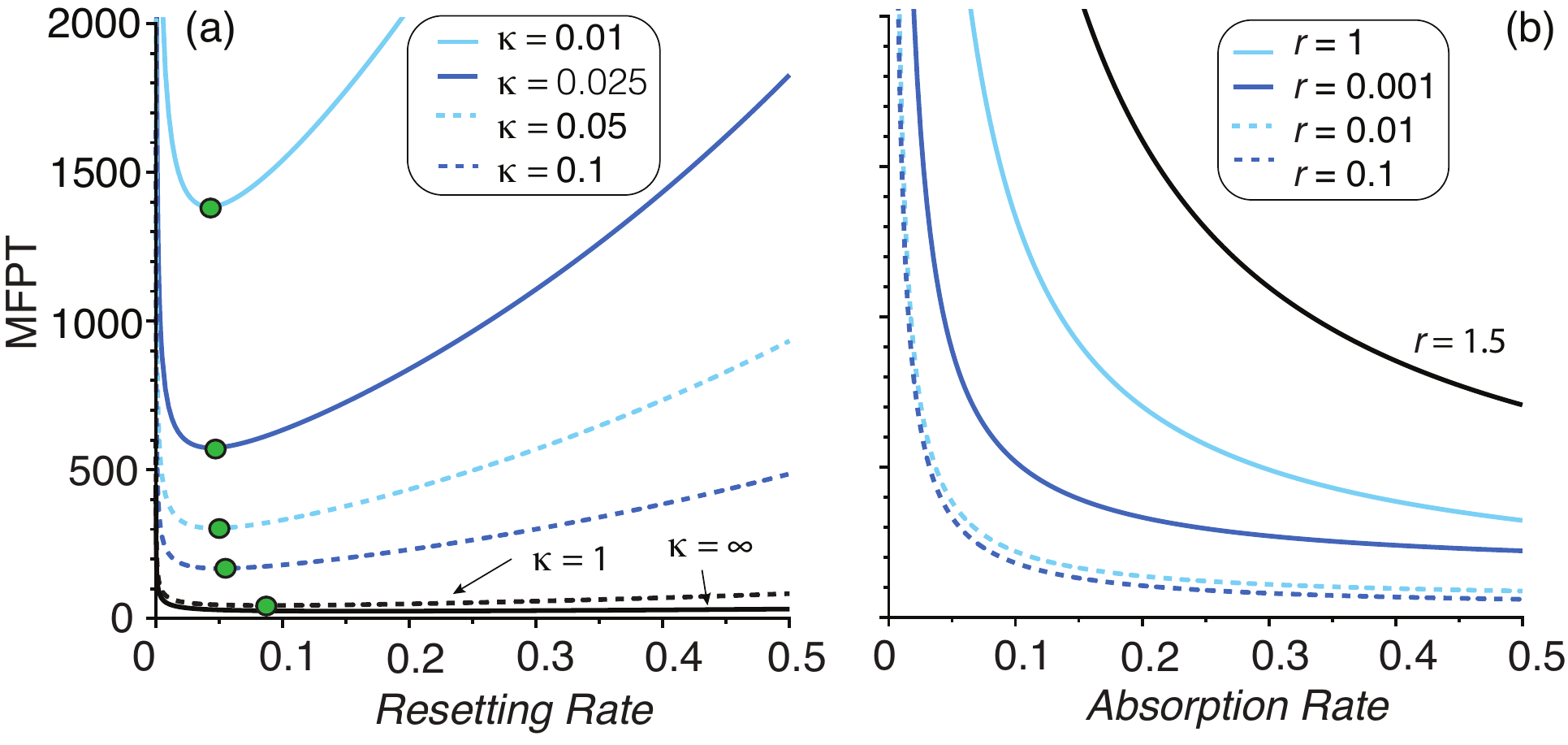}
    \caption{Single target in $\R$. (a) Plot of MFPT $T_r(x_0)$ vs the resetting rate $r$ for various $\kappa$. (b) Plot of MFPT $T_r(x_0)$ vs the absorption rate $\kappa$ for various $r$. Other parameter values are $L=1$, $x_0 = 5$, and $D = 1$. Filled circles denote $r_{\rm opt}$.}
    \label{fig2}
\end{figure}

Now allow the particle to reset to the starting position $x_0$ before being absorbed at a random time chosen according to the exponential probability density $\psi(t) = re^{-rt}$ where $r$ is the resetting rate. For this system $\pi_{r}(x_0)=1$ and the MFPT is given by equation (\ref{Ttot}):
\begin{eqnarray}
\label{TtotR}
T_{r}(\x_0)  &=\frac{1-\widetilde{J}(\x_0,r)}
{  r\widetilde{J}(\x_0,r)}.
\end{eqnarray}
with
\begin{eqnarray}
    \widetilde{J}(x_0, r) 
    &= \kappa\int_{\mathcal{U}_0}\p_1(x, r|x_0)dx= \kappa \int_{-L}^L \left(C(r)e^{\beta(r) x} + F(r)e^{-\beta(r) x}\right)dx \nonumber\\
    &= \kappa\frac{\left(e^{2\beta(r) L} - 1\right)e^{-\alpha(r)(x_0 - L)}}{\left[(\alpha(r) + \beta(r))e^{2\beta(r) L} + \alpha(r)- \beta(r)\right]\beta(r) D}.
\end{eqnarray}
First note that in the limit $\kappa \rightarrow 0$,
\begin{eqnarray}
\label{kap0}
T_r(\x_0)\approx \frac{1}{\kappa}\frac{\e^{\sqrt{r/D}x_0}}{\sinh(\sqrt{r/D}L)}
\end{eqnarray}
Clearly $T_r(x_0) \rightarrow \infty$ as $\kappa \rightarrow 0$ since there is then no absorbing target. On the other hand, $\beta(r)\rightarrow \sqrt{\kappa/D} $ as $\kappa \rightarrow \infty$ so that $T_r(\x_0)\rightarrow T_r^{\infty}(\x_0)$ with
\begin{eqnarray}
\label{kapi}
    T^{\infty}_{r}(x_0) = \frac{1}{r}\left[e^{\sqrt{r/D}(x_0 - L)} - 1\right].
\end{eqnarray}
This is the well known expression for the MFPT with resetting for a particle searching on the interval $(L,\infty)$ with an absorbing point target at $x=L$ \cite{Evans11a,Evans11b}. 
For $0 < \kappa <\infty$ $T_r$ has the exact solution
\begin{eqnarray}
\fl &T_{r}(x_0) = \frac {D\beta\left(\alpha+\beta\right)^{2}{{e}^{
 \left( x_0 - L \right) \alpha+4\beta L}}- D\beta\left(\beta - \alpha \right) ^{2}{{e}^{-\alpha\left(x_0 - L\right) }}}{ \left( 
 \left( \alpha+\beta \right) {{e}^{2\beta L}}+\beta-\alpha
 \right) \kappa r \left( {{e}^{2\beta L}}-1 \right) }\nonumber \\
\fl    &\hspace{3cm}-\frac {
 \left( -2{{e}^{2\beta L}}\alpha+ \left( \alpha+\beta
 \right) {{e}^{4\beta L}}-\beta+\alpha \right) \kappa}{ \left( 
 \left( \alpha+\beta \right) {{e}^{2\beta L}}+\beta-\alpha
 \right) \kappa r \left( {{e}^{2\beta L}}-1 \right) }.
\end{eqnarray}

\begin{figure}[t!]
    \raggedleft
    \includegraphics[width=8cm]{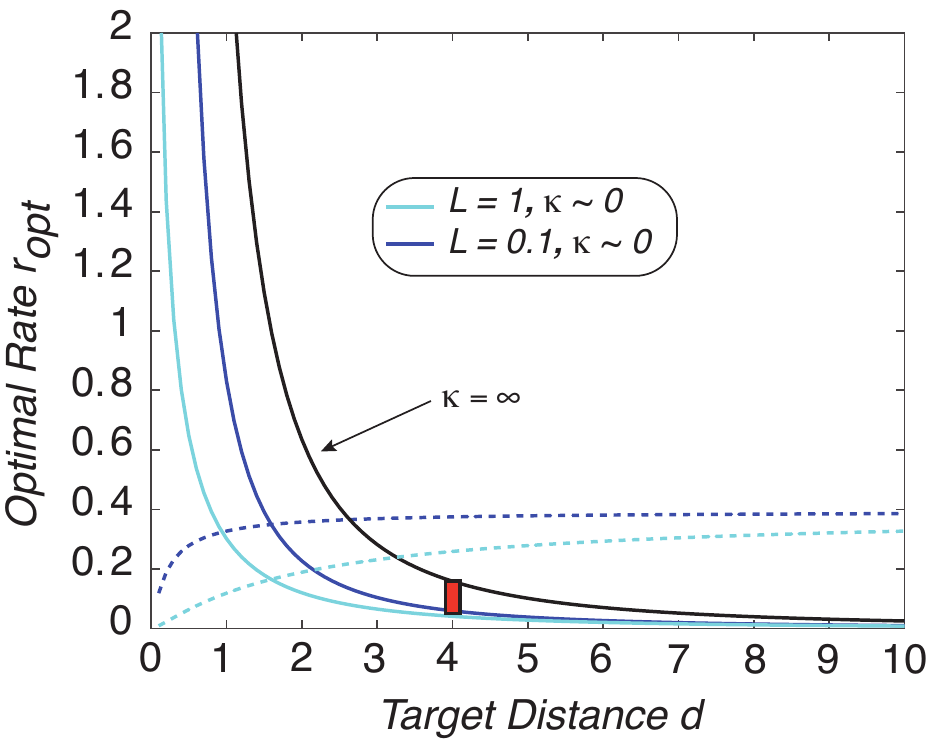}
    \caption{Plot of optimal resetting-rate $r_{\rm opt}$ as a function of target distance $d=x_0-L$ for the small and large $\kappa$ limits. Dashed curves show the ratio $r_{\rm opt}^0/r_{\rm opt}^{\infty}$ for $L=0.1,1$, which both asymptote to the value $1/\gamma^2$. Note that $r_{\rm opt}^{\infty}$ is independent of $L$. The filled rectangle indicates the range of $r_{\rm opt}$ for the parameter values used in Fig. \ref{fig2}.}
    \label{fig3}
\end{figure}

In Fig. \ref{fig2} we plot the MFPT $T_r(x_0)$ as a function of (a) the resetting rate for various $\kappa$ and (b) the absorption rate $\kappa$ for different values of the $r$. It can be seen that $T_r(x_0)\rightarrow \infty$ as $r\rightarrow 0$ since the MFPT for pure 1D diffusion is infinite. Moreover, $T_r(x_0)\rightarrow \infty$ when $r\rightarrow \infty$ since the particle resets so often it never has a chance to reach the target domain. Consistent with many other examples of stochastic resetting in which $T_r(x_0)=\infty$ when $r=0$, we find that $T_r(x_0)$ is a unimodal function of $r$ with a unique minimum at an optimal rate $r_{\rm opt}$ such that
\begin{equation}
\left .\frac{dT_r(\x_0)}{dr}\right |_{r=r_{\rm opt}}=0
\end{equation}
Using equation (\ref{kapi}) one finds that for $\kappa=\infty$ (total absorption) \cite{Evans20} 
\begin{equation}
(x_0-L)\sqrt{r_{\rm opt}^{\infty}/D}=\gamma \approx 1.5936,
\end{equation}
where $\gamma$ is the solution to the transcendental equation $ {\gamma}/{2}=1-\e^{-\gamma}$. Thus the optimal resetting rate occurs when the ratio of the distance to the target, $d=x_0-L$, to the typical distance diffused between successive resets is $\gamma$. A very different result holds for small $\kappa$; equation (\ref{kap0}) yields the transcendental equation
\begin{equation}
\tanh(\sqrt{r_{\rm opt}^0/D}L)=\frac{L}{x_0}=\frac{L}{d+L},
\end{equation}
which means that $r_{\rm opt}^0$ also depends on the target size $L$ as well as the target distance $d$. Note, in particular, that for small target sizes, $L\ll d$,
\begin{equation}
r_{\rm opt}^0\sim \frac{D}{d^2}.
\end{equation}
These two limiting cases are illustrated in Fig. \ref{fig3} where we plot $r_{\rm opt}^0$ and $r_{\rm opt}^{\infty}$ against the target distance $d$.

\setcounter{equation}{0}
\section{Spherical target in $\R^d$}
As our second example, 
consider a spherical partially absorbing target $\mathcal{U}_1 = \{{\bf x} \in \mathbb{R}^d:\|{\bf x}\| \leq R\}$ for $d=2,3$. The case $d=2$ is shown in  Fig. \ref{fig1}(b). Following \cite{Redner01}, we take the initial position of the search particle to be randomly chosen from the surface of the sphere of radius $\rho_0$.  That is,
\begin{eqnarray}
    p({\bf x}, 0|{\bf x}_0) = \frac{1}{\Omega_d \rho_0^{d - 1}} \delta(\rho - \rho_0),
\end{eqnarray}
where $\rho = \|{\bf x}\|$ and $\Omega_d$ is the surface area of a unit sphere in $\mathbb{R}^d$.  For a spherically symmetric function, equations (\ref{mastera}) and (\ref{masterb}) in Laplace space become
\numparts
\begin{eqnarray}
\label{spha}
 \fl   &\frac{\partial^2\p_1}{\partial \rho^2} + \frac{d - 1}{\rho}\frac{\partial \p_1}{\partial \rho} - \beta^2\p_1(\rho, s|\rho_0) = 0, \ \rho \in (0, R),\\ 
 \fl   &\frac{\partial^2\q}{\partial \rho^2} + \frac{d - 1}{\rho}\frac{\partial \q}{\partial \rho} - \alpha^2\q(\rho, s|\rho_0) = -\frac{1}{4\pi D \rho_0}\delta(\rho - \rho_0), \ \rho \in (R, \infty)
    \label{sphb}
\end{eqnarray}
\endnumparts
with the matching conditions
\begin{eqnarray}
    \p_1(R, s|\rho_0) = \q(R, s|\rho_0) \ \makebox{and} \ \left.\frac{\partial \q}{\partial \rho}\right|_{\rho=R} = \left.\frac{\partial \p_1}{\partial \rho}\right|_{\rho=R}.    
\end{eqnarray}
where $\alpha$ and $\beta$ are defined as they were before. As detailed in \cite{Redner01}, equations of the form (\ref{spha}) and (\ref{sphb}) can be solved in terms of modified Bessel functions:
\begin{eqnarray}
\label{pir}
    \p_1(\rho, s|\rho_0) = A \rho^\nu I_\nu(\beta \rho) + B \rho^\nu K_\nu(\beta \rho), \ \rho \in (0, R),
\end{eqnarray}
where $A, B \in \mathbb{C}$ and $\nu = 1 - d/2$.  The solution for equation (\ref{sphb}) takes the form
\begin{eqnarray}
\label{qir}
    \q(\rho, s|\rho_0) = \q_h(\rho, s|\rho_0) + G(\rho, s; \rho_0), \ \rho \in (R, \infty),
\end{eqnarray}
where $\q_h$ is the solution to the homogeneous equation satisfying $\q_h(\infty, s) = 0$ and $\q_h(R, s) = \p_1(R, s|\rho_0)$ and $G$ is the Green's function satisfying homogeneous boundary conditions at $\rho = R$ and $\rho = \infty$.  The Green' function is given by
\begin{eqnarray}
  \fl  G(\rho, s; \rho_0) = \frac{\alpha^2(\rho\rho_0)^\nu K_\nu(\alpha\rho_>)}{s\Omega_dK_\nu(\alpha R)}\left[I_\nu(\alpha\rho_<)K_\nu(\alpha R) - I_\nu(\alpha R)K_\nu(\alpha \rho_<)\right],
\end{eqnarray}
where $\rho_< = \min{(\rho, \rho_0)}$ and $\rho_> = \max{(\rho, \rho_0)}$ and the homogeneous solution is given by
\begin{eqnarray}
    \q_h(\rho, s|\rho_0) = C \rho^\nu K_\nu(\alpha \rho).
\end{eqnarray}

\subsection{Diffusion in $\mathbb{R}^2$}

For the $d=2$ case, we set $B=0$ so that $\p_1$ remains finite as $\rho \to 0$.  Applying the matching conditions and solving for $A$ and $C$ yields
\begin{eqnarray}
   &A(s) = \frac{\alpha(s)^2K_0(\alpha(s)\rho_0)}{2\pi Rs \Omega(s)},\quad
   &C(s) = \frac{\alpha(s)^2I_0(\beta(s)R)K_0(\alpha(s)\rho_0)}{2\pi RsK_0(\alpha(s)R)\Omega(s)},
\end{eqnarray}
where
\begin{eqnarray}
    \Omega(s) = \alpha(s)I_0(\beta(s)R)K_1(\alpha(s)R) + \beta(s)I_1(\beta(s)R)K_0(\alpha(s)R).
\end{eqnarray}
Therefore, the hitting probability without resetting is given by
\begin{eqnarray}
 \fl    \pi(\rho_0) = \kappa\int_{\mathcal{U}_0}\p(\rho, 0|\rho_0)d^2{\bf x} = \frac{\beta(0)}{RI_1(\beta(0)R)}\int_0^R\rho I_0(\beta(0)\rho)d\rho = 1
\end{eqnarray}
and the Laplace transform of the survival probability is
\begin{eqnarray}
\fl    \widetilde{Q}(\rho_0, s) = \frac{1}{s}\left[1 - \kappa \int_{\mathcal{U}_0}\p(\rho, s|\rho_0)d^2{\bf x}\right] = \frac{1}{s}\left[1 - \frac{2\pi R\kappa A(s)}{\beta(s)}I_1(\beta(s)R)\right].
\end{eqnarray}
The MFPT with resetting can be written as
\begin{eqnarray}
\label{T2D}
    T_{r}(\rho_0) = \frac{\widetilde{Q}(\rho_0, r)}{1 - r\widetilde{Q}(\rho_0, r)} = \frac{\beta(r) - 2\pi R\kappa A(r)I_1(\beta(r)R)}{2\pi R\kappa A(r)rI_1(\beta(r)R)}.
\end{eqnarray}

\begin{figure}[t!]
    \raggedleft
    \includegraphics[width=13cm]{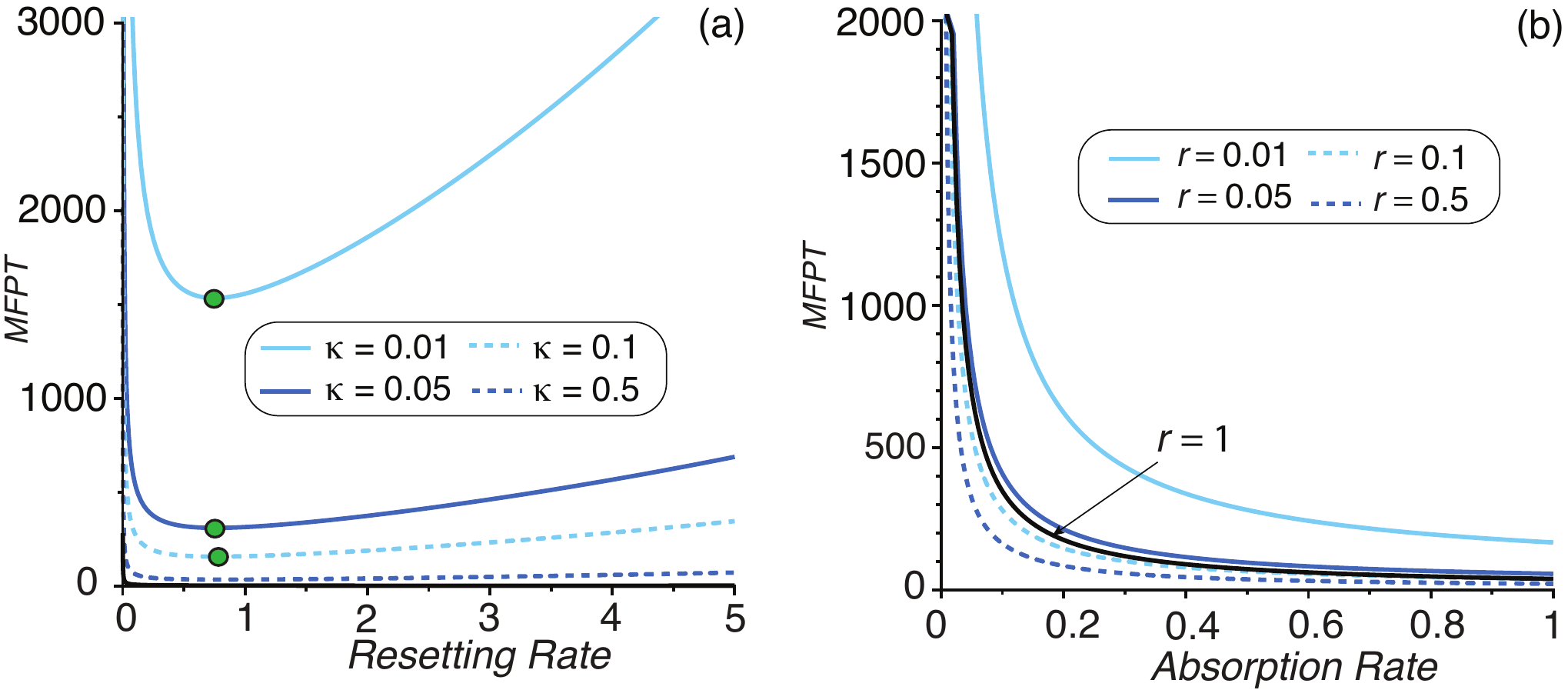}
    \caption{Spherical target in $\R^2$. (a) MFPT $T_r(\rho_0)$ vs resetting rate with $\rho_0=2$, $R=1$ and $D=1$. (b) MFPT vs absorption rate with $\rho_0=2$, $R=1$ and $D=1$. Filled circles denote $r_{\rm opt}$}
    \label{fig:sphere}
\end{figure}

\begin{figure}[b!]
    \raggedleft
    \includegraphics[width=8cm]{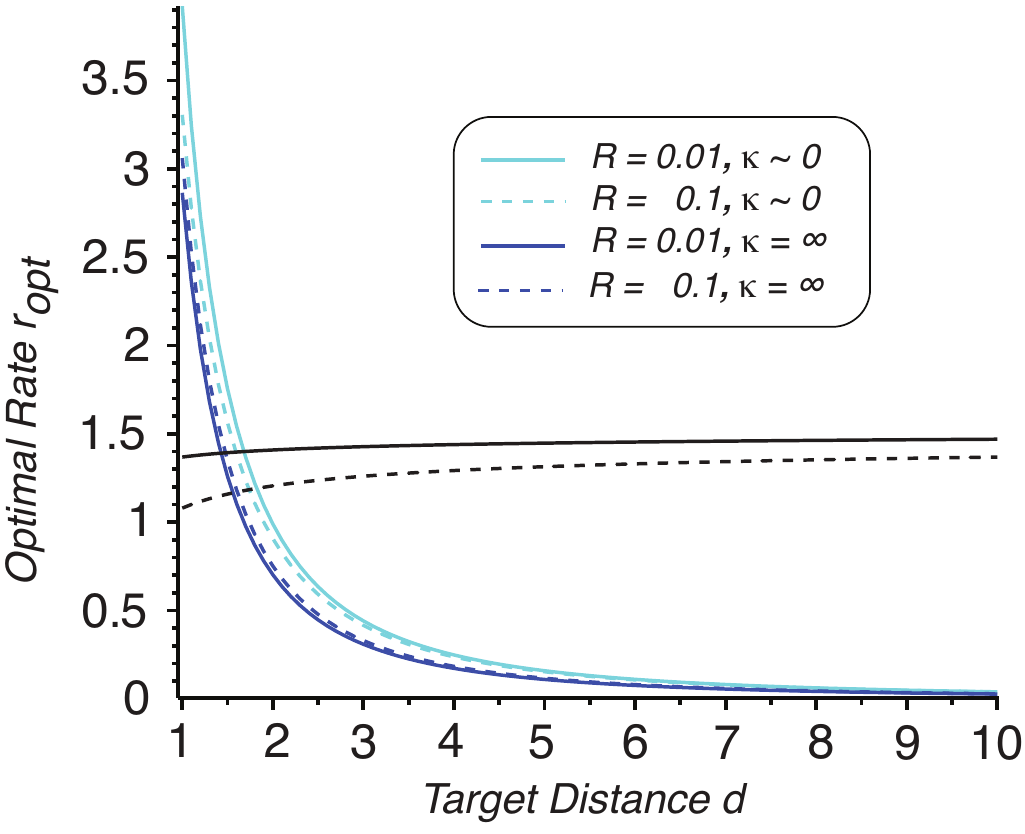}
    \caption{Plot of optimal resetting-rate $r_{\rm opt}$ in $\R^2$ as a function of target distance $d=\rho_0 - R$ for the small and large $\kappa$ limits. Pair of horizontal curves show the ratio $r_{\rm opt}^0/r_{\rm opt}^{\infty}$ for $R=0.1$ (dashed) and $R=0.01$ (solid).}
    \label{fig5}
\end{figure}

In Fig. \ref{fig:sphere} we show plots for the 2D sphere that are analogous to those of Fig. \ref{fig2} for the finite interval. Again, in the limit $\kappa\rightarrow \infty$, our results reduce to the expected MFPT for diffusion to a spherical target that is totally absorbing \cite{Evans14}:
\begin{eqnarray}
\label{goo}
    T_\infty(\rho_0) = \lim_{\kappa\to\infty} T_{r}(\rho_0) = \frac{K_0(\alpha(r)R) - K_0(\alpha(r)\rho_0)}{rK_0(\alpha(r)\rho_0)},
\end{eqnarray}
where we used the fact that $I_0(x)K_1(x) + I_1(x)K_0(x) = 1/x$.  On the other hand, in the small $\kappa$ limit, we find
\begin{equation}
     T_r(\rho_0) \approx \frac{1}{\kappa}\frac{1}{\alpha(r)RI_1\left(\alpha(r)R\right)K_0\left(\alpha(r)\rho_0\right)}.
\end{equation}
It follows that the optimal resetting rate satisfies the equation
\begin{equation}
    \fl \frac{\rho_0}{R}K_1\left(\rho_0\sqrt{r^0_{\mbox{opt}}/D}\right)I_1\left(R\sqrt{r^0_{\mbox{opt}}/D}\right) = K_0\left(\rho_0\sqrt{r^0_{\mbox{opt}}/D}\right)I_0\left(R\sqrt{r^0_{\mbox{opt}}/D}\right). 
\end{equation}
Similarly, from equation (\ref{goo}), the optimal resetting rate when $\kappa \to \infty$ is given by
\begin{eqnarray}
    \fl r^\infty_{\mbox{opt}} RK_0\left(\rho_0\sqrt{r^\infty_{\mbox{opt}}/D}\right)K_1\left(R\sqrt{r^\infty_{\mbox{opt}}/D}\right) \nonumber \\
    \fl  \quad + 2\sqrt{r^\infty_{\mbox{opt}}/D}K_0\left(\rho_0\sqrt{r^\infty_{\mbox{opt}}/D}\right)K_0\left(R\sqrt{r^\infty_{\mbox{opt}}/D}\right) \nonumber \\ 
    \fl = r^\infty_{\mbox{opt}}\rho_0K_1\left(\rho_0\sqrt{r^\infty_{\mbox{opt}}/D}\right)K_0\left(R\sqrt{r^\infty_{\mbox{opt}}/D}\right) + 2\sqrt{r^\infty_{\mbox{opt}}D}K_0\left(\rho_0\sqrt{r^\infty_{\mbox{opt}}/D}\right)^2,
\end{eqnarray}
which can be solved numerically. Using the following asymptotic approximations 
\numparts
\begin{eqnarray}
    &K_{0,1}(z) \sim \left(\frac{\pi}{2z}\right)^{1/2}e^{-z}, \ z \to  \infty\\
    &I_0(z) \sim 1 + \frac{z^2}{4}, \ z \to 0, \\
    &I_1(z) \sim \frac{z}{2}, \ z \to 0,
\end{eqnarray}
\endnumparts
we obtain for a small target size and $R \ll \rho_0$, 
\begin{eqnarray}
  &r_{\mbox{opt}}^0 \sim \frac{4D}{(d + R)^2}.
\end{eqnarray}
where $d = \rho_0 - R$. In Fig. \ref{fig5} we plot $r_{\rm opt}$ as a function of target distance $d=\rho_0 - R$ for both the small and large $\kappa$ limits. Comparison with Fig. \ref{fig3} for a 1D target shows some major differences. First, $r_{\rm opt}$ is much less sensitive to the absorption rate $\kappa$ and the target size $R$. Second, the optimal resetting rate is larger in the small $\kappa$ limit than in the large $\kappa$ limit, which is opposite to the result in 1D. (Note, however, that we recover the 1D behavior for sufficiently small distances $d$ (not shown).) Finally, the $r^0_{\mbox{opt}}$ approximation is larger by a factor of 4 compared to 1D. This is possibly explained by the increased degrees of freedom for the movement of the search particle in $\mathbb{R}^2$.

\subsection{Diffusion in $\mathbb{R}^3$}
\begin{figure}[t!]
   \raggedleft
    \includegraphics[width=13cm]{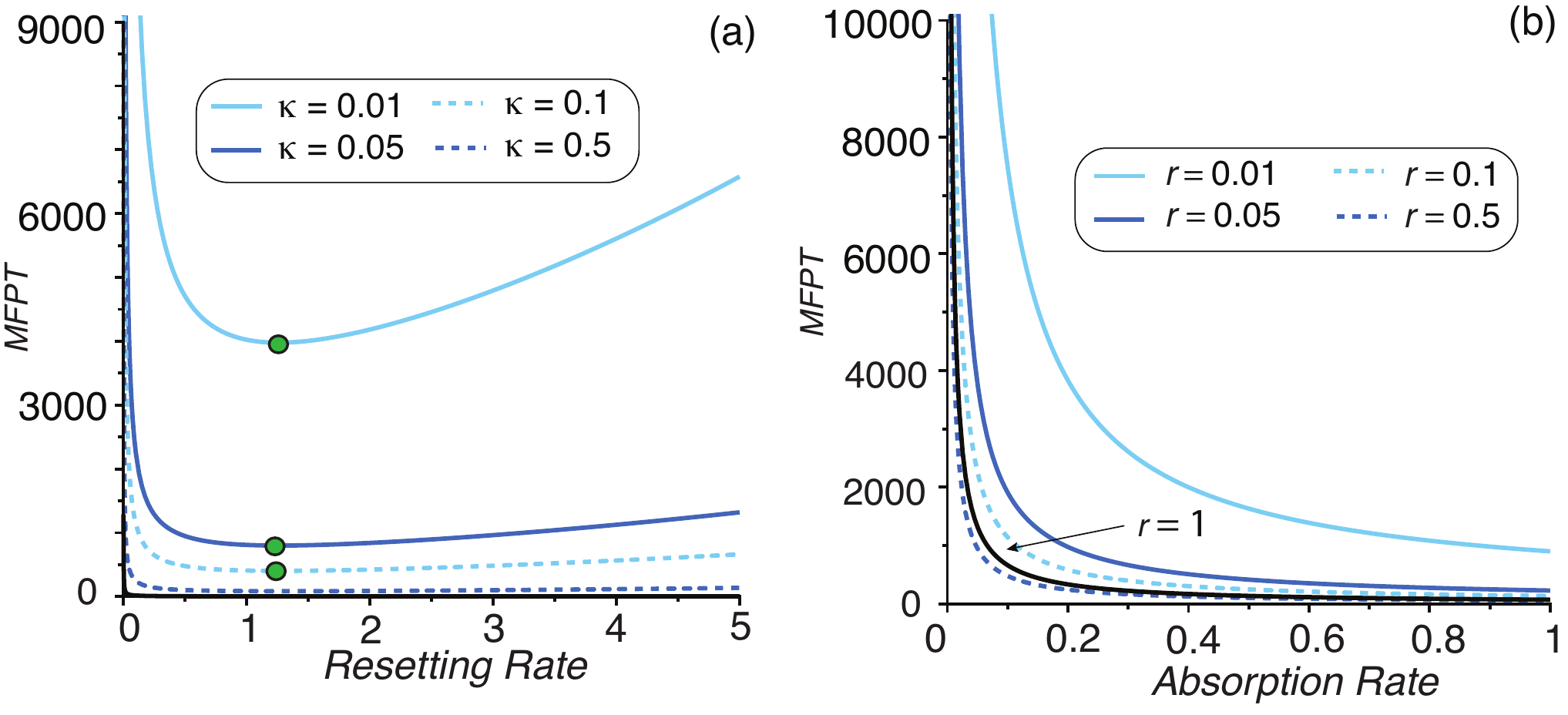}
    \caption{Spherical target in $\R^3$. (a) MFPT $T_r(\rho_0)$ vs resetting rate with $\rho_0=2$, $R=1$ and $D=1$. (b) MFPT vs absorption rate with $\rho_0=2$, $R=1$ and $D=1$. Filled circles denote $r_{\rm opt}$}
    \label{fig6}
\end{figure}

When $d=3$, $\nu = -1/2$. Therefore, we have
\numparts
\begin{eqnarray}
    &I_\nu(z) = I_{-1/2}(z) = \sqrt{\frac{2}{\pi z}}\cosh{(z)},\\
    &K_\nu(z) = K_{-1/2}(z) = \sqrt{\frac{2}{\pi z}}e^{-z},\\
    &I_{-\nu}(z) = I_{1/2}(z) = \sqrt{\frac{2}{\pi z}}\sinh{(z)}.
\end{eqnarray}
\endnumparts
Using the fact that
\begin{equation*}
    K_{-1/2}(z) = \frac{\pi}{2}\left[I_{-1/2}(z) - I_{1/2}(z)\right],
\end{equation*}
we can write the finite solution in the target region as
\begin{equation*}
    \p_1(\rho, s; \rho_0) = \frac{A}{\rho}\sinh(\beta\rho),
\end{equation*}
and the solution outside the target region as 
\begin{equation}
    \q(\rho, s; \rho_0) = \frac{C}{\rho}e^{-\alpha\rho} + \frac{\alpha e^{-\alpha(\rho_> - \rho_<)}}{8\pi s\rho\rho_0}\left[1 - e^{-2\alpha(\rho_< - R)}\right].
\end{equation}
Applying the match conditions and solving for $A$ and $C$ yields
\numparts
\begin{eqnarray}
  &A(s) = \frac{\alpha(s)^2e^{-\alpha(s)(\rho_0 - R)}}{4\pi\rho_0 s\left[\alpha(s)\sinh(\beta(s) R) + \beta(s)\cosh(\beta(s) R)\right]},\\
  &B(s) = \frac{\alpha(s)^2e^{-\alpha(s)(\rho_0 - 2R)}\sinh(\beta(s) R)}{4\pi\rho_0 s\left[\alpha(s)\sinh(\beta(s) R) + \beta(s)\cosh(\beta(s) R)\right]}.
\end{eqnarray}
\endnumparts
Therefore, the hitting probability without resetting is given by
\begin{equation}
    \fl \pi_1(\rho_0) = 4\pi\kappa\int_0^R\p(\rho, 0;\rho_0)\rho^2d\rho = \frac{R}{\rho_0} - \sqrt{\frac{D}{\kappa\rho_0^2}}\tanh\left(\sqrt{\frac{\kappa}{D}}R\right).
\end{equation}
We see that unlike the problems in $\mathbb{R}$ and $\mathbb{R}^2$, the survival probability at $t=\infty$ is nonzero due to the fact that 3-D random walks are transient. As expected, the hitting probability increases with $\kappa$ and $R$ but decreases as the starting distance between the search particle and target region increases.  In the limit $\kappa \to \infty$, we find $\pi_1(\rho_0) \to R/\rho_0$, implying that the probability of absorption is dependent only on the geometry of the problem in this limit.  The Laplace transform of the survival probability without resetting is given by
\begin{equation}
    \widetilde{Q}(\rho, s; \rho_0) = \frac{1}{s} + \frac{4\pi\kappa A(s)}{s\beta(s)^2}\left[\sinh(\beta(s) R) - \beta R\cosh(\beta(s) R)\right].
\end{equation}
Therefore, the MFPT with resetting is 
\begin{eqnarray}
    T_r(\rho_0) &= \frac{\widetilde{Q}(\rho_0, r)}{1 - r\widetilde{Q}(\rho_0, r)} \nonumber\\
    &= \frac{\beta(r)^2 + 4\pi\kappa A(r)\left[\sinh(\beta(r)R) - \beta(r)R\cosh(\beta(r)R)\right]}{4\pi\kappa rA(r)\left[\beta(r)R\cosh(\beta(r)R) - \sinh(\beta(r)R)\right]}.
\end{eqnarray}
The dependence of $T_r(\x_0)$ on $\kappa$ and $r$ is very similar to the 2D case, as shown in Fig. \ref{fig6}.

\begin{figure}[b!]
   \raggedleft
    \includegraphics[width=8cm]{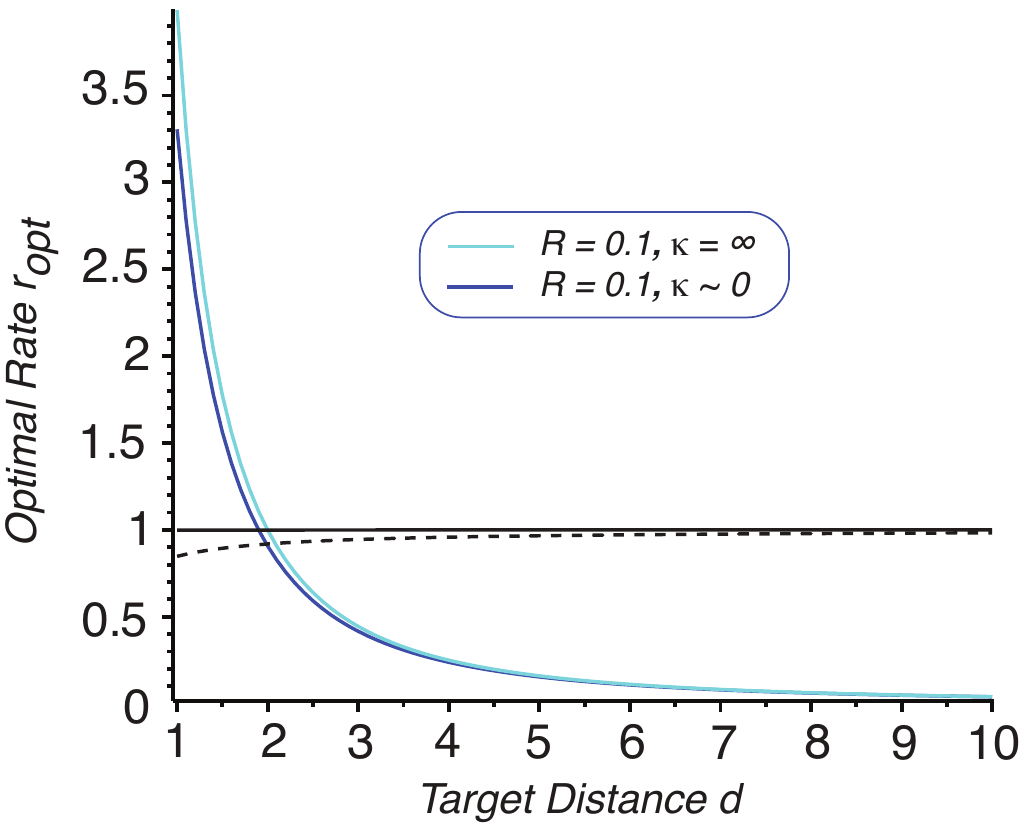}
    \caption{Plot of optimal resetting-rate $r_{\rm opt}$ in $\R^3$ as a function of target distance $d=\rho_0 - R$ for the small and large $\kappa$ limits. Pair of horizontal curves show the ratio $r_{\rm opt}^0/r_{\rm opt}^{\infty}$ for $R=0.1$ (dashed) and $R=0.01$ (solid).}
    \label{fig7}
\end{figure}

For large $\kappa$, we have \cite{Evans14}
\begin{equation}
\label{gogoo}
    T^\infty_r(\rho_0) = \lim_{\kappa \to \infty}T(\rho_0) = \frac{1}{r}\left[\frac{\rho_0}{R}e^{\sqrt{r/D}(\rho_0 - R)} - 1\right],
\end{equation}
while for small $\kappa$ we can write
\begin{equation}
    T^0_r(\rho_0) \approx \frac{\rho_0\alpha(r)e^{\alpha(r)\rho_0}}{\kappa\left[\alpha(r)R\cosh(\alpha(r)R) - \sinh(\alpha(r)R)\right]}.
\end{equation}
The optimal resetting rate in the infinite $\kappa$ limit can be obtained numerically by differentiating equation (\ref{gogoo}) with respect to $r$.
For small $\kappa$, the optimal resetting rate is determined by the equation
\begin{eqnarray}
   R&\left(\rho_0\sqrt{D}r^0_{\mbox{opt}} + D(r^0_{\mbox{opt}})^{1/2}\right)\cosh\left(\sqrt{\frac{r^0_{\mbox{opt}}}{D}}R\right) \\
   &= \left(R^2\sqrt{D}r^0_{\mbox{opt}} + D\rho_0(r^0_{\mbox{opt}})^{1/2} + D^{3/2}\right)\sinh\left(\sqrt{\frac{r^0_{\mbox{opt}}}{D}}R\right).
\end{eqnarray}
Using the approximations 
\numparts
\begin{eqnarray}
  &\cosh(z) \sim 1 + \frac{z^2}{2}, \ z \to 0,\\
  &\sinh(z) \sim z + \frac{z^3}{6}, \ z \to 0,\\
  &\sqrt{1 - z^2} \sim 1 - \frac{z^2}{2} - \frac{z^4}{8}, \ z \to 0,
\end{eqnarray}
\endnumparts
we find
\begin{equation}
    r^0_{\mbox{opt}} \sim \frac{4D}{(d + R)^2}.
\end{equation}
Hence, in the small $\kappa$ limit, the optimal resetting rate for a spherical target in $\mathbb{R}^3$ depends on the starting distance and target size in the same way as a spherical target in $\mathbb{R}^2$. Fig. \ref{fig7} shows the behavior of the optimal resetting rate as a function of the starting distance, which shows that there is an even weaker dependence on $\kappa$ compared to 2D, with a slight increase of $r_{\rm opt}$ for larger $\kappa$.

\section{Pair of absorbing targets in $\R^2$}

\subsection{Splitting probabilities and MFPTs for multiple targets with resetting}
So far we have focused on a single target. However, it is possible to extend the analysis to multiple targets along the lines of \cite{Chechkin18,Bressloff20A}. Let $\calU_k\subset \R^d$ denote the $k$-th partially absorbing target, $k=1,\ldots,N$, and set $\calU_a=\bigcup_{k=1}^N\calU_k$. Equations (\ref{mastera})--(\ref{masterc}) become
\numparts 
\begin{eqnarray}
\label{master2a}
	\frac{\partial q(\x,t|\x_0)}{\partial t} &=& D\nabla^2 q(\x,t|\x_0), \ \x\in \R^d \backslash \calU_a,\\
	\label{master2b}
	\frac{\partial p_k(\x,t|\x_0)}{\partial t} &=& D\nabla^2 p_k(\x,t|\x_0) -\kappa p_k(\x,t|\x_0),\ \x\in \calU_k,
	\end{eqnarray}
together with the continuity conditions 
\begin{equation}
\label{master2c}
\fl q(\x,t|\x_0)=p_k(\x,t|\x_0),\quad \nabla q(\x,t|\x_0)\cdot \n =  \nabla p_k(\x,t|\x_0)\cdot \n\quad \x \in \partial \calU_k.
\end{equation}
\endnumparts
The probability flux into the $k$-th target at time $t$ is 
\begin{eqnarray}
\label{Jm}
	J_k(\x_0,t)&= \kappa\int_{\calU_k} p_k(\x,t|\x_0)d\x,\ k = 1,\ldots,N.
	\end{eqnarray}
Hence, the splitting probability that the particle is eventually captured by the $k$-th target is
\begin{equation}
\label{split2}
\pi_k(\x_0)=\int_0^{\infty}J_k(\x_0,t')dt' =\widetilde{J}_k(\x_0,s),
\end{equation}
where $\widetilde{J}_k(\x_0,s)$ denotes the Laplace transform of $J_k(\x_0,t)$.
The survival probability that the particle hasn't been absorbed by a target in the time interval $[0,t]$, having started at $\x_0$ is now given by
\begin{equation}
\label{QN}
\fl Q(\x_0,t)=\int_{\R^d}p(\x,t|\x_0)d\x=\int_{\R^d\backslash \calU_a}q(\x,t|\x_0)d\x+\sum_{k=1}^N\int_{\calU_k}p_k(\x,t|\x_0)d\x.
\end{equation}
Following along similar lines to the single target case, we find that
\begin{equation}
\label{QNL}
s\widetilde{Q}(\x_0,s)-1=- \sum_{k= 1}^N \widetilde{J}_k(\x_0,s).
\end{equation}
and
\begin{equation}
\label{splitN0}
\sum_{k=1}^N\pi_k(\x_0)=1-Q_{\infty}(\x_0).
\end{equation}
It will also convenient to introduce the probability that the particle is captured by the $k$-th target after time $t$:
\begin{equation}\Pi_k(\x_0,t)=\int_t^{\infty}J_k(\x_0,t')dt'.
\end{equation}
In Laplace space
\begin{equation}
s\widetilde{\Pi}_k(\x_0,s)-\pi_k=-\widetilde{J}_k(\x_0,s).
\end{equation}

Now suppose that we include stochastic resetting to an initial position $\x_0\notin \calU_a$. Let the discrete random variable $K(t)\in \{0,1,\ldots,N\}$ indicate whether the particle has been absorbed by the $k$-th target ($K(t)=k \neq 0$) or has not been absorbed by any target ($K(t)=0$) in the time interval $[0,t]$.
The FPT that the particle is absorbed by the $k$-th target is then
\begin{equation}
\calT_{r,k}=\inf\{t>0; \X(t)\in \calU_k,\ K(t)=k\},
\end{equation}
with $\calT_{r,k}=\infty$ if the particle is absorbed by another target.
The following results then hold \cite{Bressloff20A}. First,
the splitting probability that the search process with resetting finds the $k$-th target is 
\begin{eqnarray}
\pi_{r,k}(\x_0)&=\frac{\pi_{k}(\x_0)-r\widetilde{\Pi}_{k}(\x_0,r)}{ 
1-r\widetilde{Q}(\x_0,r)} =\frac{\widetilde{J}_k(\x_0,r)}{ 
\sum_{j=1}^N\widetilde{J}_j(\x_0,r)}.
\label{Piee}
\end{eqnarray}

Similarly, the Laplace transformed conditional FPT density for the $k$th target is
\begin{eqnarray}
\label{Tcond1}
 \pi_{r,k}(\x_0)\widetilde{f}_{r,k}(\x_0,s)&=\frac{\pi_{k}(\x_0)-(r+s)\widetilde{\Pi}_{k}(\x_0,r+s)}{1-r\widetilde{Q}(\x_0,r+s)}.
\end{eqnarray}
The Laplace transform of the FPT density is the moment generator of the conditional FPT ${\mathcal T}_k$:
\begin{equation}
\pi_{r,k}T_{r,k}^{(n)}=\E[{\mathcal T}_k^n1_{\Omega_k}]=\left .\left (-\frac{d}{ds}\right )^n\E[\e^{-s{\mathcal T}_k}1_{\Omega_k}]\right |_{s=0}.
\end{equation}
For example, the conditional MFPT $T_{r,k}=T_{r,k}^{(1)}$ is
\begin{eqnarray}
\label{Tcond2}
\fl \pi_{r,k}(\x_0)T_{r,k}(\x_0)  &=\frac{\widetilde{\Pi}_{k}(\x_0,r) +r\widetilde{\Pi}_{k}'(\x_0,r)}
{1-
r\widetilde{Q}(\x_0,r)}+\left [\frac{\pi_{k}(\x_0)-r\widetilde{\Pi}_{k}(\x_0,r)}{ 
1-r\widetilde{Q}(\x_0,r)}\right ] \left [\frac{-r\widetilde{Q}'(\x_0, r) }{ 
1-r\widetilde{Q}(\x_0,r)}\right ],\nonumber
\end{eqnarray}
where $'$ denotes differentiation with respect to $r$. Using equation (\ref{Piee}) and the fact that
\begin{eqnarray*}
    \widetilde{Q}(\x_0, r) = \frac{Q_\infty(\x_0)}{r} +\sum_{k=1}^N\widetilde{\Pi}_k(\x_0, s),    
\end{eqnarray*}
we obtain the following expression for the conditional MFPTs :
\begin{eqnarray}
\label{Tcond3}
\fl & \pi_{r,k}(\x_0)T_{r,k}(\x_0)  =\frac{\widetilde{\Pi}_{k}(\x_0,r) +r\widetilde{\Pi}_{k}'(\x_0,r)}
{1-
r\widetilde{Q}(\x_0,r)}+\pi_{r,k}(\x_0)\left [\frac{Q_\infty(\x_0)/r-r\sum_k\widetilde{\Pi}_k'(\x_0,r)}{ 
1-r\widetilde{Q}(\x_0,r)}\right ].\nonumber 
\end{eqnarray}
Finally, summing over $k$ yields the unconditional MFPT
\begin{eqnarray}
\label{Ttotg}
\fl T_{r}(\x_0)  &:=\sum_{k=1}^N \pi_{r,k}(\x_0)T_{r,k}(\x_0) =\frac{\widetilde{Q}(\x_0,r)}{1-
r\widetilde{Q}(\x_0,r)} =\frac{1-  \sum_{k= 1}^N \widetilde{J}_k(\x_0,r)}
{  r\sum_{j=1}^N\widetilde{J}_j(\x_0,r)}.
\end{eqnarray}

\subsection{Partially absorbing disc and annulus}

\begin{figure}[t!]
\raggedleft
\includegraphics[width=8cm]{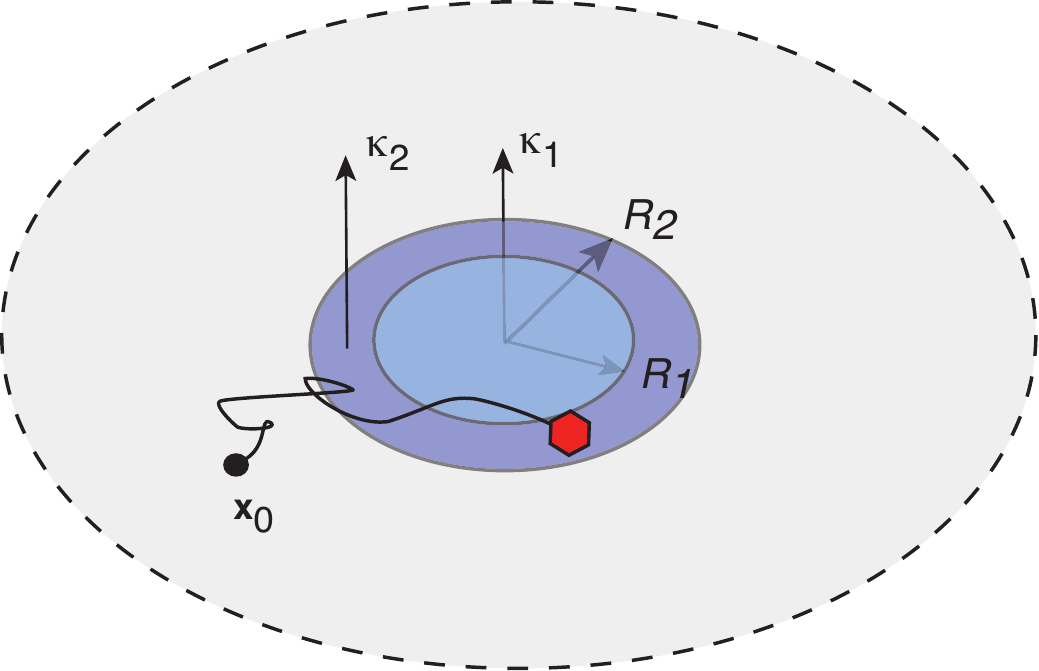} 
\caption{Pair of radially symmetric targets: a disc of radius $R_1$ and an annular region $R_1\leq \rho \leq R_2$.}
\label{fig8}
\end{figure}

Let us now consider two concentric spheres centered at the origin with radii $R_1$ and $R_2$, respectively, such that $R_1 < R_2$.  The spheres define two partially absorbing target regions: $\mathcal{U}_1 = \{\x \in \mathbb{R}^2:\|\x\|<R_1\}$ and $\mathcal{U}_2 = \{\x \in \mathbb{R}^2:R_1<\|\x\|<R_2\}$ with absorption rates $\kappa_1$ and $\kappa_2$ respectively, see Fig. \ref{fig8}. Note that in the limit $\kappa_2\rightarrow \infty$ the system reduces to a single totally absorbing target of radius $R_2$, whereas in the limit $\kappa_1\rightarrow \infty$ with $\kappa_2=0$ we have a totally absorbing target of  radius $R_1$. In addition, the MFPT is greatest in the limiting case where $\kappa_1 \to 0$ and $\kappa_2$ is finite but the MFPT is smallest in the paradigm where the annulus to become totally absorbing. The Laplace transform of the probability density for a diffusing particle in this region satisfies
\numparts
\begin{eqnarray}
 \fl   &\frac{\partial^2\p_1}{\partial \rho^2} + \frac{1}{\rho}\frac{\partial \p_1}{\partial \rho} - \beta_1(s)^2\p_1(\rho, s|\rho_0) = 0, \ \rho \in (0, R_1),\\ 
 \fl   &\frac{\partial^2\p_2}{\partial \rho^2} + \frac{1}{\rho}\frac{\partial \p_2}{\partial \rho} - \beta_2(s)^2\p_2(\rho, s|\rho_0) = 0, \ \rho \in (R_1, R_2),\\ 
\fl    &\frac{\partial^2\q}{\partial \rho^2} + \frac{1}{\rho}\frac{\partial \q}{\partial \rho} - \alpha(s)^2\q(\rho, s|\rho_0) = -\frac{1}{4\pi D\rho_0}\delta(\rho - \rho_0), \ \rho \in (R_2, \infty),
\end{eqnarray}
\endnumparts
and matching conditions at the boundaries $\rho = R_1$ and $\rho = R_2$ where $\beta_1(s) = \sqrt{(s+\kappa_1)/D}$ and $\beta_2(s) = \sqrt{(s + \kappa_2)/D}$. The general solution is of the form
\numparts
\begin{eqnarray}
\fl    &\p_1(\rho, s|\rho_0) = A(s)\rho^\nu I_\nu(\beta_1(s)\rho) + B(s)\rho^\nu K_\nu(\beta_1(s)\rho), \ \rho \in (0, R_1),\\
\fl    &\p_2(\rho, s|\rho_0) = C(s)\rho^\nu I_\nu(\beta_2(s)\rho) + E(s)\rho^\nu K_\nu(\beta_2(s)\rho), \ \rho \in (R_1, R_2),\\
 \fl   &\q(\rho, s|\rho_0) = P(s)\rho^\nu K_\nu(\alpha(s)\rho) + R(s)\rho^\nu I_\nu(\alpha(s)\rho), \ \rho \in (R_2, \infty). 
\end{eqnarray}
\endnumparts
Next, we set $B=0$ so that the probability density is bounded at $\rho=0$. Imposing continuity of the Laplace transforms of the probability density and flux at $\rho = R_1$ and $\rho = R_2$ gives a four-by-four matrix equation with a unique solution for the coefficients $A$, $C$, $E$, and $F$.  We find that
\begin{eqnarray}
   \widetilde{J}_1(\rho_0, s) &= 2\kappa_1\pi\int_0^{R_1}\p_1(\rho,s|\rho_0)\rho d\rho = \frac{2\pi \kappa_1R_1A(s)}{\beta(s)}I_1(\beta_1(s)R_1)
\end{eqnarray}
and
\begin{eqnarray}
\fl  &  \widetilde{J}_2(\rho_0, s) 
    = 2\kappa_2\pi\int_{R_1}^{R_2}\p_2(\rho,s|\rho_0)\rho d\rho= \frac{2\pi\kappa_2}{\beta_2(s)}\bigg [C(s)\left(R_2I_1(\beta_2(s)R_2) - R_1I_1(\beta_2(s)R_1\right)\nonumber \\
 \fl   & \hspace{5cm} - E(s)\left(R_2K_1(\beta_2(s)R_2) - R_1K_1(\beta_2(s)R_1\right)\bigg ].
\end{eqnarray}
The splitting probabilities without resetting are given by
\begin{eqnarray}
    \pi_j(\rho_0) = \widetilde{J}_j(\rho_0, 0), \ j = 1,2.
\end{eqnarray}
The corresponding splitting probabilities with resetting are given by equation (\ref{Piee})
and the unconditioned MFPT with resetting is given by equation (\ref{Ttot}).  Note that 
\begin{eqnarray}
    T_\infty(\rho_0) = \lim_{\kappa_2\to\infty}T_r(\rho_0) = \frac{K_0(\alpha(r)R_2) - K_0(\alpha(r)\rho_0)}{K_0(\alpha(r)\rho_0)}.
\end{eqnarray}
That is, if we allow the annulus to become totally absorbing, we recover the result for a single totally absorbing target of radius $R_2$ in $\mathbb{R}^2$. (Recall, that the initial position of the particle is taken to be outside  the outer sphere.)

\begin{figure}[t!]
       \raggedleft
    \includegraphics[width=13cm]{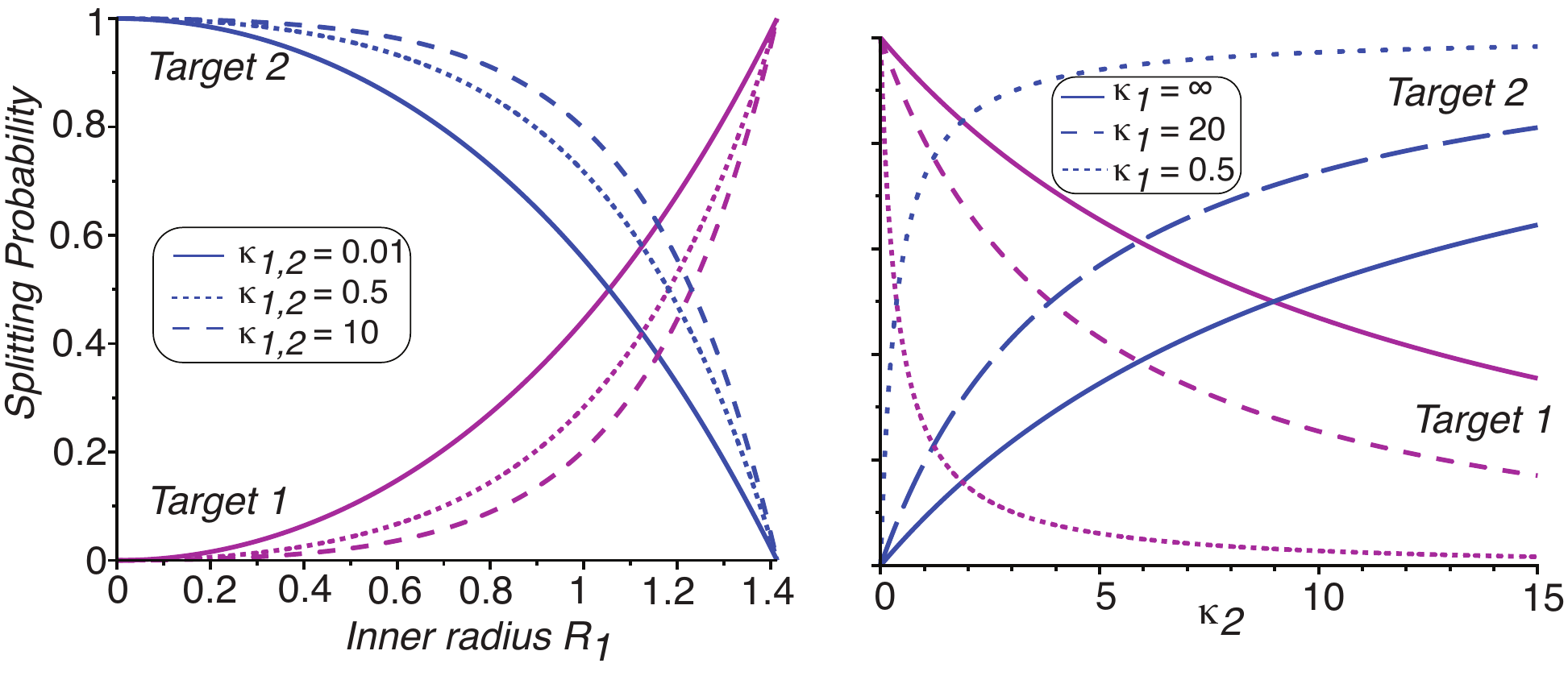}
    \caption{Annulus (target 2) and inner sphere (target 1) in $\R^2$. (a) Splitting probabilities with resetting vs spherical target radius $R_1$ for differenet absorption rates $\kappa_1=\kappa_2$. (b) Splitting probabilities with resetting vs annulus absorption rate $\kappa_2$ for different $\kappa_1$. Other parameter values are $\rho_0=3$, $R_2=\sqrt{2}$, $D=1$, and $r=1$}
    \label{fig9}
\end{figure}

\begin{figure}[b!]
      \raggedleft
    \includegraphics[width=13cm]{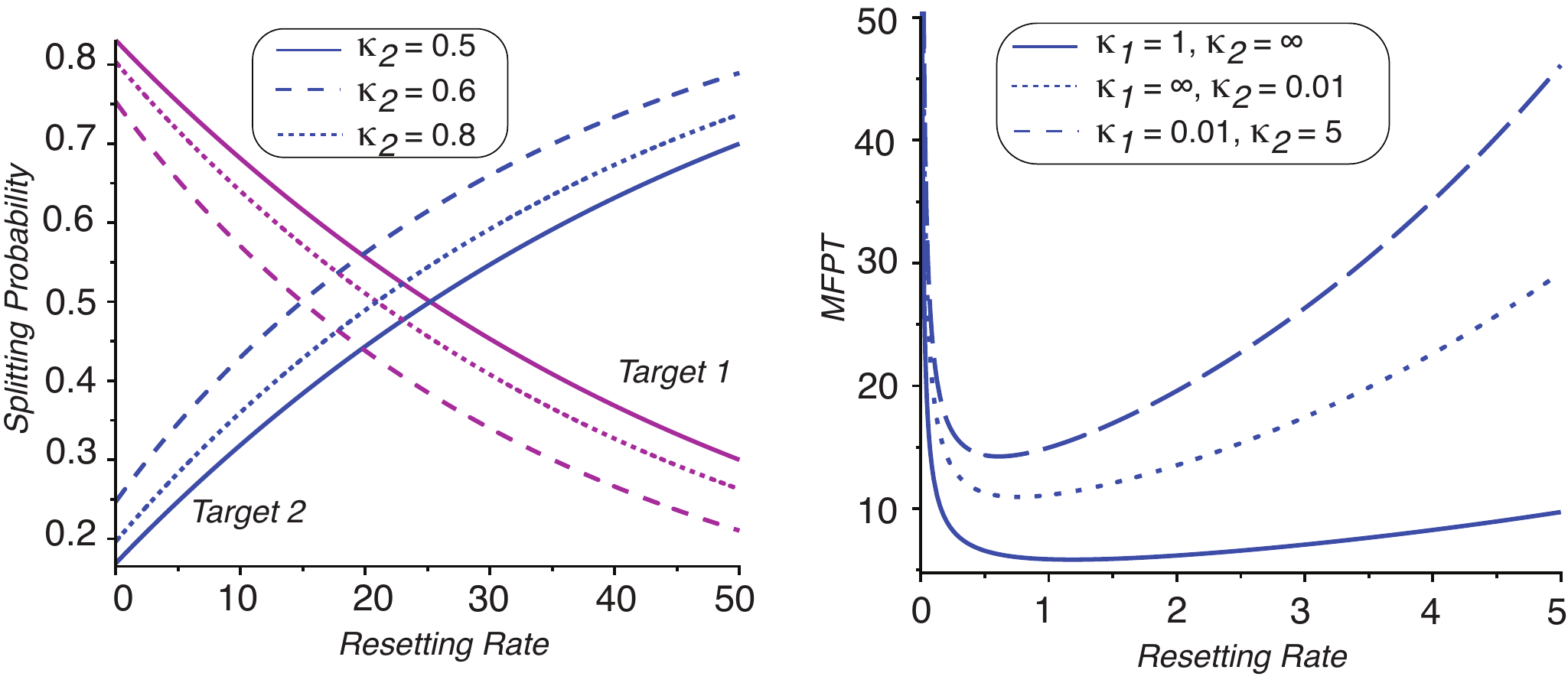}
     \caption{Annulus and spherical targets in $\R^2$. (a) Splitting probabilities with resetting vs resetting rate $r$ with $\rho_0=3$, $R=1$, $R_2=\sqrt{2}$, and $D=1$. (b) Unconditioned MFPT with resetting vs resetting rate for different limiting cases with $\rho_0=3$, $R=1$, $R_2=\sqrt{2}$, $D=1$, and $r=1$}
     \label{fig10}
\end{figure}

In Fig. \ref{fig9} we plot the splitting probabilities of the annular and spherical targets as a function of various combinations of absorptions rates $\kappa_{1,2}$ and inner radius $R_1$. Note that there is a crossover phenomenon, whereby the inner target becomes more likely to absorb the particle than the annulus when $R_1$ and $\kappa_1$ are sufficiently large. In addition, if $\kappa_1=\kappa_2=\kappa$, say, then the crossover radius $R_1$ increases with $\kappa$. On the other hand, the annular region can have a higher splitting probability even when the inner target becomes totally absorbing ($\kappa_1=\infty$). In Fig. \ref{fig10}(a) we plot the splitting probabilities as a function of the resetting rate $r$. Choosing a parameter regime in which $\pi_{r,1} > \pi_{r,2}$ without resetting ($r\rightarrow 0$), we find that there exists a crossover point $r^*$ such that $\pi_{r^*,1} = \pi_{r^*,2}$ with $r^*$ a non-monotonic function of $\kappa_2$ for fixed $\kappa_1$.  Finally, in Fig. \ref{fig10}(b), we plot $T_r$ as a function $r$ for different combinations of $\kappa_1$ and $\kappa_2$. Qualitatively, we see that $T_r$ depends on $r$ in a similar fashion to the previous geometries. 

\section{Extended kinetic scheme}

So far we have assumed that a particle with a target is directly absorbed at a rate $\kappa$. A natural extension of the basic model is to consider a more general chemical kinetic scheme for absorption. In particular, suppose that on entering a target, the particle binds to the target and has to undergo a sequence of reversible reactions before being absorbed. Let $S_0$ denote the free particle and $S_m$, $m=1,\ldots,M$, one of the bound states. Consider the reaction scheme
\begin{equation}
S_0\Markov{\alpha_1}{\gamma_0} S_1 \Markov{\alpha_2}{\gamma_1}S_2\cdots  \Markov{\alpha_M}{\gamma_{M-1}}S_M\overset{\gamma_M}\rightarrow \emptyset.
\end{equation}
 Let $C_{m}(t)$ denote the probability that the particle is in state $m$ within the target, and set 
 \begin{equation}
 J(\x_0,t)=\gamma_0\int_{\calU_1}p_1(\x,t|\x_0)d\x. 
 \end{equation}
 (For the sake of illustration, we focus on a single target.) Equation (\ref{masterb}) is then replaced by the system of equations
\numparts
\begin{eqnarray}
\label{kinetica}
	\frac{\partial p_1}{\partial t} &= D\nabla^2 p_1-\gamma_0 p_1+\frac{\alpha_1}{|\calU_1|}C_{1},\ \x\in \calU_1,\\
	\label{kineticb}
	\frac{dC_{1}}{dt}&=  J-(\alpha_1+\gamma_{1})C_{1}+\alpha_{2}C_{2},\\
	\frac{dC_{m}}{dt}&= \gamma _{m-1} C_{m-1}-(\alpha_m+\gamma_{m})C_{m}+\alpha_{m+1}C_{m+1}
	\end{eqnarray}
	for $m=2,\ldots,M-1$, and
	\begin{eqnarray}
	\frac{dC_{M}}{dt}&= \gamma _{M-1} C_{M-1}-(\alpha_M+\gamma_M)C_{M}.
	\label{kineticc}
\end{eqnarray}
\endnumparts
We now have to modify the definition of the survival probability in equation (\ref{Q1}) according to
\begin{eqnarray}
\label{Q1K}
 Q(\x_0,t)
 &=\int_{\R^d\backslash \calU_1}q(\x,t|\x_0)d\x+ \int_{\calU_1}p_1(\x,t|\x_0)d\x+\sum_{m=1}^M C_{m}(t).
\end{eqnarray}
Differentiating both sides of this equation with respect to $t$ and using equations (\ref{mastera}) and (\ref{kinetica}) implies that 
\begin{eqnarray}
 \frac{\partial Q(\x_0,t)}{\partial t}
 &=- \int_{\partial \calU_1}\nabla q\cdot \n d\sigma+  \int_{\partial \calU_1}\nabla p_1\cdot \n d\sigma \nonumber \\
 &\quad -\gamma_0  \int_{ \calU_1}p_1(\x,t|\x_0)d\x+  {\alpha_1} C_{1}(t) + \sum_{m=1}^{M}\frac{dC_{m,}}{dt}.
\end{eqnarray}
Imposing equation (\ref{masterc}) and summing equations (\ref{kineticb})--(\ref{kineticc}) then gives
\begin{eqnarray}
 \frac{\partial Q(\x_0,t)}{\partial t}
 &= -\gamma_M  C_{M}(t).
\label{Q2K}
\end{eqnarray}
Laplace transforming equation (\ref{Q2K}) and imposing the initial condition $Q(\x_0,0)=1$ gives
\begin{equation}
\label{QLK}
s\widetilde{Q}(\x_0,s)-1=- \gamma_M  \C_{M}(s).
\end{equation}

Laplace transforming equations (\ref{kinetica})--(\ref{kineticc}), assuming that the particle is initially outside any of the targets, we have
 \numparts 
\begin{eqnarray}
\label{kineticLTa}
	&0= D\nabla^2 \p_1 -(\gamma_0+s) \p_1+\frac{\alpha_1}{|\calU_1|}\C_{1},\ \x\in \calU_1,
	\\
	&s\C_{m} = \widetilde{J} \delta_{m,1} +\sum_{m'=1}^M\Gamma_{mm'}\C_{m'}
	\label{kineticLTb}
\end{eqnarray}
\endnumparts
for $m=1,\ldots,M$,
where $\Gamma_{mm'}$ is an element of the tridiagonal matrix
\begin{equation}
\fl {\bm \Gamma}=\left (\begin{array}{cccccc} -\alpha_1-\gamma_1 & \alpha_2 &0 &0 &\ldots&0 \\ \gamma_1&-\alpha_2-\gamma_2 & \alpha_3 &0 & \ldots&0
\\0& \gamma_2 &-\alpha_3-\gamma_3 & \alpha_4&\ldots&0\\
\vdots & \vdots & \vdots & \vdots &\vdots  &\vdots \\
0 & 0 &\ldots & 0 & \gamma_{M-1} & -\alpha_M -\kappa
\end{array} \right ).
\end{equation}
It follows that
\begin{equation}
\label{Cm}
\C_{m}=  (s{\bf I}-{\bm \Gamma})^{-1}_{m1}\widetilde{J}.
\end{equation}
Finally, substituting (\ref{Cm}) with $m=1$ into equation (\ref{kineticLTa}) gives
\begin{equation}
\label{extkin}
0= D\nabla^2 \p_1 - (\gamma_0+s) \p_1+\frac{\alpha_1}{|\calU_1|}(s{\bf I}-{\bm \Gamma})^{-1}_{11}\widetilde{J},\ \x\in \calU_1.
\end{equation}

\subsection{Spherical target in $\R^d$ and $M=1$}
In the case of a spherical target in $\R^d$, $d=2,3$, and a single-step reaction, equation (\ref{spha}) becomes
\begin{eqnarray}
 \fl   &D\left [\frac{\partial^2\p_1}{\partial \rho^2} + \frac{d - 1}{\rho}\frac{\partial \p_1}{\partial \rho}\right ] -(\gamma_0+s)\p_1(\rho, s|\rho_0) +\frac{\gamma_0\alpha_1}{s+\alpha_1+\gamma_1}\langle \p_1\rangle= 0,    
\end{eqnarray}
for $ \rho \in (0, R)$, where
\begin{equation}
\label{aloo}
\langle \p_1\rangle =\frac{1}{|\calU_1|} \int_{\calU_1}\p_1(\x,s|\x_0)d\x=\frac{d}{R^{d}}\int_0^R \p_1(\rho,s|\rho_0)\rho^{d-1}d\rho.
\end{equation}
In addition, equation (\ref{Cm}) simplifies to
\begin{equation}
\C_{1}=\frac{ \widetilde{J}}{s+\alpha_1+\gamma_1}.
\end{equation}
Introducing the change of variables
\begin{equation}
\label{cvar}
P=\p_1-c_1(s)\langle \p_1\rangle,\quad c_1(s)=\frac{\gamma_0}{\gamma_0+s}\frac{\alpha_1}{s+\alpha_1+\gamma_1},
\end{equation}
we see that 
\begin{eqnarray}
\label{eqP}
  & \frac{1}{\rho^{d-1}}\frac{d}{d\rho} \rho^{d-1} \frac{d P}{d \rho} -\frac{\gamma_0+s}{D}P= 0,    
\end{eqnarray}
which has a solution of the form (\ref{pir}):
\begin{eqnarray}
\label{pir2}
   P= A \rho^\nu I_\nu(\beta_0 \rho) + B \rho^\nu K_\nu(\beta_0 \rho), \ \rho \in (0, R), \ \beta_0=\sqrt{\frac{s+\gamma_0}{D}},
\end{eqnarray}
The solution outside the target is still given by equation (\ref{qir}),
 \begin{eqnarray}
\label{qir2}
\q(\rho, s|\rho_0) = C \rho^\nu K_\nu(\alpha \rho) + G(\rho, s; \rho_0), \ \rho \in (R, \infty),
\end{eqnarray}
and the matching conditions become (after dropping the explicit dependence on $\rho_0$)
\begin{eqnarray}
    P(R, s ) +c_1(s)\langle \p_1\rangle= \q(R, s ) \ \makebox{and} \ \left.\frac{\partial \q}{\partial \rho}\right|_{\rho=R} = \left.\frac{\partial P}{\partial \rho}\right|_{\rho=R}.    
\end{eqnarray}
It remains to determine the unknown term $\langle \p_1\rangle$. Substituting equations (\ref{cvar}) and (\ref{pir2}) into (\ref{aloo}) and setting $P=P(\rho,s)$ yields the self-consistency condition 
\begin{eqnarray}
(1-c_1(s))\langle \p_1\rangle&=\frac{d}{R^d} \int_{0}^R P(\rho,s)\rho^{d-1} d\rho .\end{eqnarray}
Multiplying both sides of equation (\ref{eqP}) by $\rho^{d-1}$ and integrating with respect to $\rho$ implies
\begin{equation}
R^{d-1}P'(R,s)= \frac{\gamma_0+s}{D} \int_{0}^R P(\rho,s)\rho^{d-1} d\rho
\end{equation}
and, hence,
\begin{eqnarray}
(1-c_1(s))\langle \p_1\rangle&=\frac{d}{R} \frac{D}{\gamma_0+s}P'(R,s) .
\end{eqnarray}
The matching conditions are now
\begin{eqnarray}
\label{BCEK2}
 \fl    P(R, s ) +\frac{c_1(s)}{1-c_1(s)}\frac{d}{R} \frac{D}{\gamma_0+s}P'(R,s)= \q(R, s ) \ \makebox{and} \ \left.\frac{\partial \q}{\partial \rho}\right|_{\rho=R} = \left.\frac{\partial P}{\partial \rho}\right|_{\rho=R}.    
\end{eqnarray}
Requiring that the solutions are finite, these matching conditions determine the unknown coefficients in $P(\rho,s)$. One can then solve for the Laplacve transformed survival probaility using (\ref{QLK}):
\begin{eqnarray}
\label{QLK2}
s\widetilde{Q}(\x_0,s)-1&=-   \gamma_1\C_{1}(s)  =-\frac{ \gamma_1\widetilde{J}(\x_0,s)}{s+\alpha_1+\gamma_1}   \nonumber \\
&=-\frac{ \gamma_1}{s+\alpha_1+\gamma_1} \frac{\gamma_0}{\gamma_0+s}\frac{2^{d-1}\pi D R^{d-1}P'(R,s)}{1-c_1(s)} .
\end{eqnarray}

\begin{figure}[t!]
    \center
    \includegraphics[width=13cm]{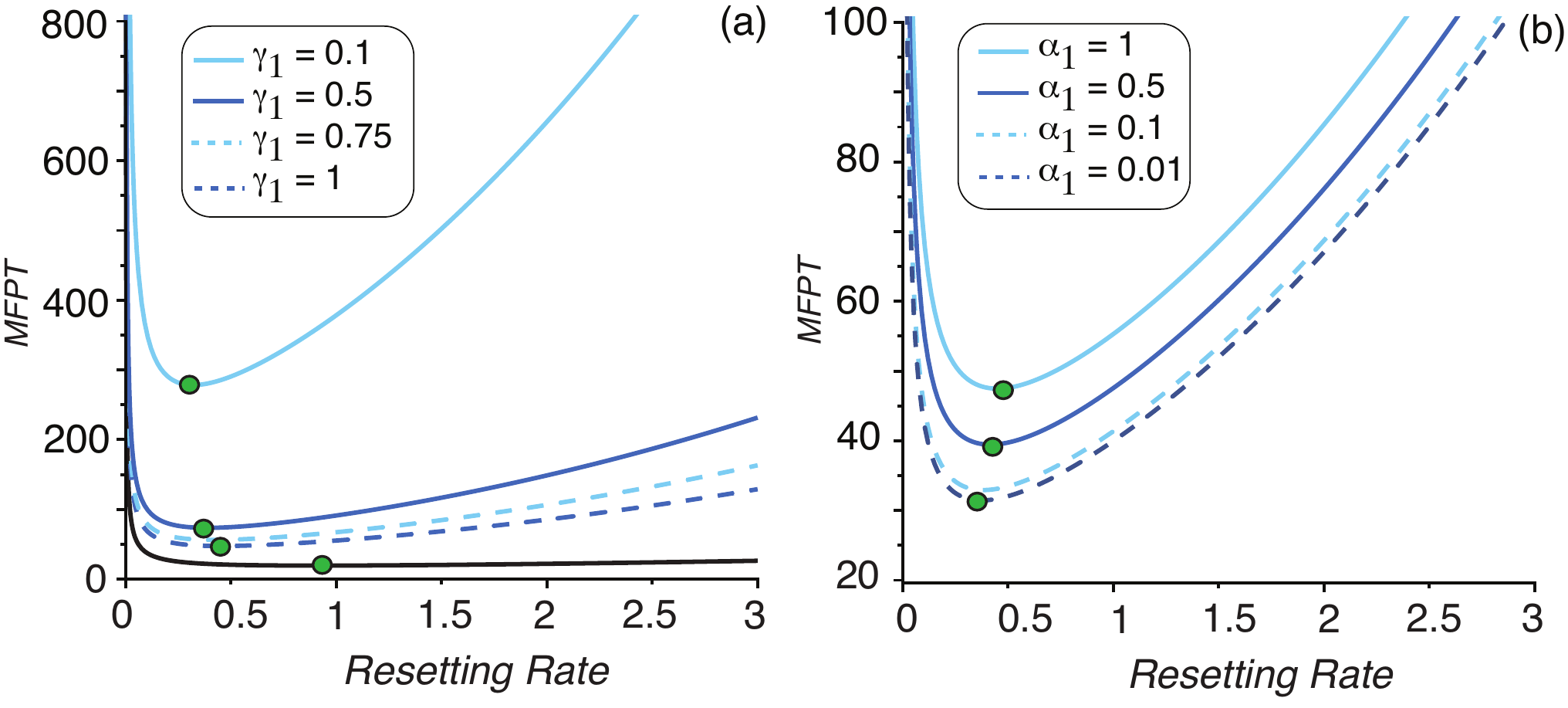}
    \caption{Spherical target in $\R^2$ with enzyme kinetic extension. (a) MFPT $T_r(\rho_0)$ vs resetting rate for various $\gamma_1$ and $\alpha_1=1$. (b) Corresponding plots for fixed $\gamma_1=1$ and various $\alpha_1$. Other parameter values are $\gamma_0=1$, $\rho_0=2$, $R=1$ and $D=1$. Filled circles indicate $r_{\rm opt}$.}
    \label{fig11}
\end{figure}

Setting $s=r$ and substituting into equation (\ref{Ttot}) gives
\begin{eqnarray}
\label{Ttotext}
 T_{r}(\x_0)  =\frac{\widetilde{Q}(\x_0,r)}{1-
r\widetilde{Q}(\x_0,r)} =\frac{r+\alpha_1+\gamma_1-  {\gamma_1}  \widetilde{J}(\x_0,r)}
{  r\gamma_1  \widetilde{J}(\x_0,r)}. 
\end{eqnarray}
Note that in the limit $\gamma_1\rightarrow \infty$ such that $c_1\rightarrow 0$, the particle is absorbed as soon as it binds to the substrate, and we recover equation (\ref{T2D}) with $\kappa_1=\gamma_0$.
For the sake of illustration, consider the case $d=2$. We then have $P = A(s)K_0\left(\beta_0\rho\right)$ and (\ref{BCEK2}) gives
\begin{eqnarray}
\fl   &A(s) = \frac{\alpha(s)^2\beta(s)[c_1(s) - 1]K_0(\alpha(s)\rho_0)}{2\pi s\left[\beta(s)^2R[c_1(s) - 1]I_1(\beta(s)R)K_0(\alpha(s)R) - 2\alpha(s)\mathcal{D}(s)K_1(\alpha(s)R)\right]},\\
  \fl  &B(s) = \frac{(\alpha(s)^2/R)D(s)K_0(\alpha(s)\rho_0)}{\pi s K_0(\alpha(s)R)\left[\beta(s)^2R[c_1(s) - 1]I_1(\beta(s)R)K_0(\alpha(s)R) - 2\alpha(s)\mathcal{D}(s)K_1(\alpha(s)R)\right]}\nonumber \\
  \fl
\end{eqnarray}
where
\begin{eqnarray*}
    \mathcal{D}(s) = c_1(s)I_1(\beta(s)R) - \frac{\beta(s)R}{2}[c_1(s) - 1]I_0(\beta(s)R).
\end{eqnarray*}
The Laplace transform of the survival probability can be written as 
\begin{equation}
\fl    \widetilde{Q}(\rho_0, s) = \frac{2\pi RD\beta(s)\gamma_0\gamma_1A(s)I_1(\beta(s)R) + (s + \alpha_1 + \gamma_1)(s + \gamma_0)(c_1(s) - 1)}{s(s + \gamma_0)(c_1(s) - 1)(s + \alpha_1 + \gamma_1)}
\end{equation}
with the MFPT given by (\ref{Ttot}). In Fig. \ref{fig11} we plot the MFPT $T_r(\x_0)$ as a function of the resetting rate for various $\gamma_1$ and $\alpha_1$. Note, in particular, that $r_{\rm opt}$ is more sensitive to variations in the absorption rate $\gamma_1$ in the extended kinetic scheme compared to the direct absorption case.

\section{Discussion}
In this paper, we developed a theoretical framework to study search processes with stochastic resetting and target regions that are partially absorbing. That is, the search particle is absorbed at a rate $\kappa$ upon entering the target region. We considered various target geometries in $\R^d$, and determined how target splitting probabilities (in the case of more than one target) and the MFPT to absorption depended on the absorption rate $\kappa$, the resetting rate $r$ and the target geometry. We also explored the parameter-dependence of the optimal resetting rate $r_{\rm opt}$ that minimizes the MFPT.

There are many possible extensions of the analysis of partially absorbing targets developed in this paper. First, as in the case of totally absorbing targets, one could consider other stochastic search processes such as active search by run-and-tumble particles \cite{Evans18}, directed intermittent search \cite{Bressloff20b}, and Levy flights \cite{Kus14}. As mentioned in section 2, the resetting protocol could also include delays such as finite return times and refractory periods \cite{Mendez19,Evans19a,Mendez19a,Bodrova20,Pal20,Bressloff20A,Evans20}. As illustrated in section 6, another major aspect of partially absorbing targets is the nature of the absorption process within each target. One interesting generalization would be to incorporate a non-Markovian reaction scheme analogous to a study of anomalous diffusion within spiny dendrites \cite{Fedotov08}. The latter example considers particles diffusing along a one-dimensional dendritic cable that is studded with partially absorbing spines. If the exchange of a particle between a spine and the parent dendrite is non-Markovian then the transport becomes subdiffusive. Finally, rather than assuming that the boundary of a target is fully permeable to a diffusing particle, one could consider a semi-permeable membrane.

\end{document}